\documentclass[longauth]{aa}
\UseRawInputEncoding 
\usepackage[utf8]{inputenc}
\usepackage{pdflscape}
\usepackage{xcolor}
\usepackage{lscape}
\usepackage{float}
 \usepackage[hidelinks]{hyperref}
   \hypersetup{
      colorlinks = true,
      citecolor ={cyan}
   }
\usepackage{longtable}
\usepackage{graphicx}
\newcommand{\paperI}{Paper~I}

\newcommand{\kms}{km\,${\rm s}^{-1}$}

\defcitealias{BLOeM-I}{Paper~I}

\usepackage{txfonts}

\begin{document}

   \title{Binarity at LOw Metallicity (BLOeM)\thanks{Based on observations collected at the European Southern Observatory under ESO program ID 112.25W2.}$^,$\thanks{Tables ... are only available in electronic form at the CDS via anonymous ftp to cdsarc.u-strasbg.fr (130.79.128.5) or via \url{http://cdsweb.u-strasbg.fr/cgi-bin/qcat?J/A+A/}.}} 
    \subtitle{The multiplicity properties and evolution of BAF-type supergiants}

   \author{L. R.~Patrick\inst{\ref{inst:cab}}
          \and D. J.~Lennon\inst{\ref{inst:iac}}
          \and F.~Najarro\inst{\ref{inst:cab}} 
          \and T.~Shenar\inst{\ref{inst:TelAv}}
          \and J.~Bodensteiner\inst{\ref{inst:eso}} 
          \and  H.~Sana\inst{\ref{inst:kul}} 
          \and  P.~A.~Crowther\inst{\ref{inst:sheffield}} 
          \and N. Britavskiy \inst{\ref{inst:rob}}
          \and N.~Langer\inst{\ref{inst:bonn}}
          \and A.~Schootemeijer\inst{\ref{inst:bonn}}
          \and C. J. Evans \inst{\ref{inst:esa_stsci}}
          \and L.~Mahy \inst{\ref{inst:rob}}
          \and Y.~G{\"o}tberg \inst{\ref{inst:ista}}
          \and S.~E.~de Mink \inst{\ref{inst:mpa}}
          \and F.~R.~N.~Schneider\inst{\ref{inst:hits},\ref{inst:ari}}
          \and A.~J.~G.~O'Grady\inst{\ref{inst:cmu}}
          \and J.~I.~Villase\~{n}or\inst{\ref{inst:mpia}}
          \and M.~Bernini-Peron\inst{\ref{inst:ari}}
          \and D.~M.~Bowman\inst{\ref{inst:newcastle}, \ref{inst:kul}}
          \and A.~de Koter \inst{\ref{inst:antonpannekoek}, \ref{inst:kul}}          
          \and K.~Deshmukh \inst{\ref{inst:kul}}
          \and A.~Gilkis \inst{\ref{inst:cambridge}}
          \and G.~Gonz\'alez-Tor\`a\inst{\ref{inst:ari}}
          \and V.~M.~Kalari\inst{\ref{inst:gemini}}
          \and Z.\~Keszthelyi\inst{\ref{inst:naoj}}
          \and I.~Mandel\inst{\ref{inst:monash},\ref{inst:ozgrav}}
          \and A.~Menon\inst{\ref{inst:columbia}}          
          \and M.~Moe\inst{\ref{inst:wyoming}} 
          \and L.~M.~Oskinova\inst{\ref{inst:up}}
          \and D.~Pauli\inst{\ref{inst:up}}
          \and M.~Renzo\inst{\ref{inst:AZ}}
          \and A.~A.~C.~Sander\inst{\ref{inst:ari}}
          \and K.~Sen\inst{\ref{inst:umk}}
          \and M.~Stoop\inst{\ref{inst:antonpannekoek}}
          \and J.~Th. van Loon\inst{\ref{inst:keele}}
          \and S.~Toonen\inst{\ref{inst:antonpannekoek}}
          \and F.~Tramper\inst{\ref{inst:cab}}
          \and J.~S. Vink\inst{\ref{inst:armagh}} 
          \and C.~Wang\inst{\ref{inst:mpa}}
     }

   \institute{  
{Centro de Astrobiolog\'ia (CSIC-INTA), Ctra.\ Torrej\'on a Ajalvir km 4, 28850 Torrej\'on de Ardoz, Spain\label{inst:cab}}\\ \email{lrpatrick@cab.inta-csic.es}    
\and
{Instituto de Astrof\'isica de Canarias, C. V\'ia L\'actea, s/n, 38205 La Laguna, Santa Cruz de Tenerife, Spain\label{inst:iac}}
\and  
{The School of Physics and Astronomy, Tel Aviv University, Tel Aviv 6997801, Israel\label{inst:TelAv}};
   \and
{ESO - European Southern Observatory, Karl-Schwarzschild-Strasse 2, 85748 Garching bei M\"unchen,
Germany \label{inst:eso}}
   \and
{Institute of Astronomy, KU Leuven, Celestijnenlaan 200D, 3001 Leuven, Belgium\label{inst:kul}}
\and
{Department of Physics \& Astronomy, Hounsfield Road, University of Sheffield, Sheffield, S3 7RH, United Kingdom\label{inst:sheffield}} 
\and {Royal Observatory of Belgium, Avenue Circulaire/Ringlaan 3, B-1180 Brussels, Belgium} \label{inst:rob}
\and
{Argelander-Institut f\"{u}r Astronomie, Universit\"{a}t Bonn, Auf dem H\"{u}gel 71, 53121 Bonn, Germany\label{inst:bonn}}
\and 
{{European Space Agency (ESA), ESA Office, Space Telescope Science Institute, 3700 San Martin Drive, Baltimore, MD 21218, USA}\label{inst:esa_stsci}}
\and
{{Institute of Science and Technology Austria (ISTA), Am Campus 1, 3400 Klosterneuburg, Austria}\label{inst:ista}}
\and {Max-Planck-Institute for Astrophysics, Karl-Schwarzschild-Strasse 1, 85748 Garching, Germany\label{inst:mpa}}
\and
{Zentrum f\"ur Astronomie der Universit\"at Heidelberg, Astronomisches Rechen-Institut, M\"onchhofstr. 12-14, 69120 Heidelberg, Germany\label{inst:ari}} 
\and 
{School of Mathematics, Statistics and Physics, Newcastle University, Newcastle upon Tyne, NE1 7RU, UK\label{inst:newcastle}}
\and
Heidelberger Institut f{\"u}r Theoretische Studien, Schloss-Wolfsbrunnenweg 35, 69118 Heidelberg, Germany\label{inst:hits}
\and 
{McWilliams Center for Cosmology \& Astrophysics, Department of Physics, Carnegie Mellon University, Pittsburgh, PA 15213, USA\label{inst:cmu}}
\and {Max-Planck-Institut f\"{u}r Astronomie, K\"{o}nigstuhl 17, D-69117 Heidelberg, Germany\label{inst:mpia}}
\and {{Anton Pannekoek Institute for Astronomy, University of Amsterdam, Science Park 904, 1098 XH Amsterdam, the Netherlands} \label{inst:antonpannekoek}}
\and {Institute of Astronomy, University of Cambridge, Madingley Road, Cambridge CB3 0HA, United Kingdom} \label{inst:cambridge}
\and {Gemini Observatory/NSF's NOIRLab, Casilla 603, La Serena, Chile}\label{inst:gemini}    
\and
{Center for Computational Astrophysics, Division of Science, National Astronomical Observatory of Japan, 2-21-1, Osawa, Mitaka, Tokyo 181-8588, Japan\label{inst:naoj}}
\and {{School of Physics and Astronomy, Monash University, Clayton VIC 3800, Australia}\label{inst:monash}}
\and
{{ARC Centre of Excellence for Gravitational-wave Discovery (OzGrav), Melbourne, Australia}\label{inst:ozgrav}}
\and
{{Department of Astronomy, Pupin Hall, Columbia University, New York City, New York 10027, U.S.A}\label{inst:columbia}}
\and {Department of Physics and Astronomy, University of Wyoming, 1000 E. University Ave., Dept. 3905, Laramie, WY 82071, USA\label{inst:wyoming}}
\and {Institut f\"ur Physik und Astronomie, Universit\"at Potsdam, Karl-Liebknecht-Str. 24/25, 14476 Potsdam, Germany\label{inst:up}}
\and {{Department of Astronomy \& Steward Observatory, 933 N. Cherry Ave., Tucson, AZ 85721, USA}\label{inst:AZ}
}
\and
{{Institute of Astronomy, Faculty of Physics, Astronomy and Informatics, Nicolaus Copernicus University, Grudziadzka 5, 87-100 Torun, Poland}\label{inst:umk}
}
\and
{Lennard-Jones Laboratories, Keele University, ST5 5BG, UK\label{inst:keele}}
\and 
{Armagh Observatory, College Hill, Armagh, BT61 9DG, Northern Ireland, UK\label{inst:armagh}}
}

   \date{Received -; accepted -}

  \abstract
  {Given the uncertain evolutionary status of blue supergiant stars, their multiplicity properties hold vital clues to better understand their origin and evolution. As part of The Binarity at LOw Metallicity (BLOeM) campaign in the Small Magellanic Cloud we present a multi-epoch spectroscopic survey of 128 supergiant stars of spectral type B5--F5, which roughly correspond to initial masses in the range 6 to 30\,M$_\odot$. 
  The observed binary fraction for the B5-9 supergiants is 25\,$\pm$\,6\,\% (10\,$\pm$4\%) and 5\,$\pm$\,2\,\% (0\%) for the A-F stars, using a radial velocity (RV) variability threshold of 5\,\kms (10\,\kms) as a criterion for binarity.
  Accounting for observational biases we find an intrinsic multiplicity fraction of less than 18\,\% for the B5-9 stars and 8$^{+9}_{-7}$\% for the AF stars, for the orbital periods up to 10$^{3.5}$\,day and mass-ratios ($q$) in the range $0.1 < q < 1$. 
  The large stellar radii of these supergiant stars prevent short orbital periods but we demonstrate that this effect alone cannot explain our results.
  We assess the spectra and RV time series of the detected binary systems and find that only a small fraction display convincing solutions.
  We conclude that the multiplicity fractions are compromised by intrinsic stellar variability such that the true multiplicity fraction may be significantly smaller.
  Our main conclusions from comparing the multiplicity properties of the B5-9 and AF supergiants to that of their less evolved counterparts is that such stars cannot be explained by a direct evolution from the main sequence.
  Furthermore, by comparing their multiplicity properties to red supergiant stars we conclude that the AF supergiant stars are neither progenitors nor descendants of red supergiants.
  }

   \titlerunning{Multiplicity of BLOeM BAF-type supergiants}
   \authorrunning{Patrick, L. R. et al.}

   \maketitle
%
%
\section{Introduction}\label{sec:intro}

The Binarity at LOw Metallicity (BLOeM) survey is a large ESO program (PI: Shenar, dPI: Bodensteiner; ID: 112.25R) that is delivering a 25-epoch spectroscopic survey of $\sim$1000 massive stars in the Small Magellanic Cloud (SMC) over a baseline of two years.  
The primary observational goals include the estimation of binary fractions in different evolutionary phases, the determination of orbital configurations of systems with periods $P \lesssim 3\,$yr, and the search for dormant black-hole binary candidates (OB+BH).
An over-arching science imperative is to enable a deep understanding of the impact of binary evolution on massive star populations and their environments. Further details of the survey and its objectives are provided in \citet[][hereafter \paperI]{BLOeM-I}, which also presents an overview of the first nine epochs of data, distributed over 30--65\,d between Sep--Dec 2023.
A preliminary variability analysis of these first epochs, focused primarily on detecting short period binary systems, is underway and concentrated on five broad cohorts of sources: 
the O-type stars \citep{BLOeM-O}; 
the early B-type dwarfs and giants covering broadly B0-B3 stars and luminosity classes (LC) V-III \citep{BLOeM-BV};
the early B-type supergiants and bright giants covering spectral types B0--B3 and LC II-I \citep[][hereafter early-BSGs]{BLOeM-BI}; 
the classical Oe and Be stars \citep{BLOeM-Be}; 
and finally the B5 to F supergiants (hereafter BAF supergiants) that are the subject of this paper. 
All subgroups (apart from the OBe stars) are illustrated in the Hertzsprung-Russell diagram (HRD) in Figure\,\ref{fig:hrd}.

The distribution of B- to F-type supergiants
in the HRD reflects a long-standing puzzle in massive star evolution. 
The observed distribution of sources from the main sequence through to the F-type supergiant regime, only modified by the upper limit of the Humphreys-Davidson (HD) limit \citep{hdlimit}, is contrary to theoretical expectations and observed in all nearby star-forming galaxies
\citep[early references include][]{hdlimit,humphreysmcelroy,humphreys1983,fitzpatrickgarmany}.
Single star evolutionary theory predicts a post main-sequence gap in the HRD. 
This is because after the end of the main sequence, stars evolve rapidly towards the core helium burning phase and appear as red supergiants (RSG) or BAF-type supergiants, with another gap between these two phases. 
This latter gap, sometimes referred to as the `Hertzsprung gap', 
is indeed observed as a dearth of G-type supergiants, see previous references, while the issue of whether or not a BAF-type supergiant phase precedes or follows the RSG phase, i.e. the presence of blue loops in the HRD and the mass range within which these occur, depends sensitively on the mixing physics employed in the models. For a general discussion of these issues in the context of the SMC/LMC see, for example, \citet{lennon2010,Schootemeijer2019,georgy2021,Schootemeijer2021}.
Complicating the picture further, most massive stars are born in binary or multiple systems that undergo interaction with a companion during their lifetime \citep{sana2012} leading to more complicated evolutionary scenarios.
In this context the stars found within the gap at the end of the main-sequence may consist of post-interaction binaries, possibly hosting stripped stars or black holes, or perhaps even the results of stellar mergers 
\citep{deMink2014,langer2020,menon2024,Goetberg2018,villasenor2023,wang2024}.

From an observational perspective, the stellar properties, in particular CNO abundances \citep{trundle2005,trundle2007,crowther2006,wessmayer2022} and pulsation properties \citep{saio2013,kraus2015,Bowman2019,arias2023,linhao2024}, are diagnostics that are used to assign evolutionary status. 
Such diagnostics are able to distinguish between core hydrogen burning (i.e. main sequence) stars with abundances dictated by mass-loss and/or mixing, core-helium burning stars whose properties are consistent with pre- or post-RSG evolution, mass transfer products in a binary system, and stellar mergers. 
However, given the importance of binary evolution in massive stars, critical information is provided by their multiplicity properties, which, when compared to other evolutionary stages, is particularly insightful to define evolutionary histories. 
For example, an absence of blue supergiants with orbital separations too short to fit a RSG star could be interpreted as evidence for blue loops, since this phase would remove all but the longest period companions~\citep{stothers1970}.
Furthermore, \cite{mcevoy2015} studied 52 B-type supergiants in the 30 Doradus region of the LMC finding that the 18 binaries in this sample are mostly located at spectral types earlier than about B2, and speculated that the hotter sample constituted main sequence stars, the cooler single stars being core helium burning (see their Figure 4). 
A single cool binary in that sample was discovered to have a stripped star primary \citep{villasenor2023}.

The BLOeM data for the BAF supergiants therefore provide a unique opportunity for a systematic multiplicity study of late-B to F-type supergiants.  
This is a demographic that is relatively unexplored, as most spectroscopic monitoring studies of this group having focused on a small number of the most luminous late-B and A-type supergiants, the $\alpha$ Cygni variables \citep[][]{partha1987,rosendhal1970,kaufer1997}.
This class of object display radial velocity (RV) variations caused by radial pulsations on time scales of days, and with typical amplitudes of 10--20\,\kms\ that are capable of masking and/or mimicking binary motion. 
\citet{burki1983} studied 45 F-type supergiants and found a binary fraction of 18\,\%, albeit all but a few were judged to have periods longer than their 5 year time baseline.


In this article we use multi-epoch spectroscopic observations taken as part of the BLOeM campaign to investigate the variability of 128 BAF supergiant stars in the SMC. 
The sample selection is described in Section~\ref{sec:obs}, and our methodology of determining multi-epoch RV time series is described in Section~\ref{sec:methods}. 
The results are presented in Section~\ref{sec:results} where we identify candidate binary systems and perform a multiplicity analysis based on the determined RV time series.
This is discussed in Section~\ref{sec:discussion} 
and our main conclusions are presented in Section~\ref{sec:conclusions}.

\section{Sample and observations}\label{sec:obs}

\begin{figure*}[htp]
\centering
\includegraphics[width=\textwidth]{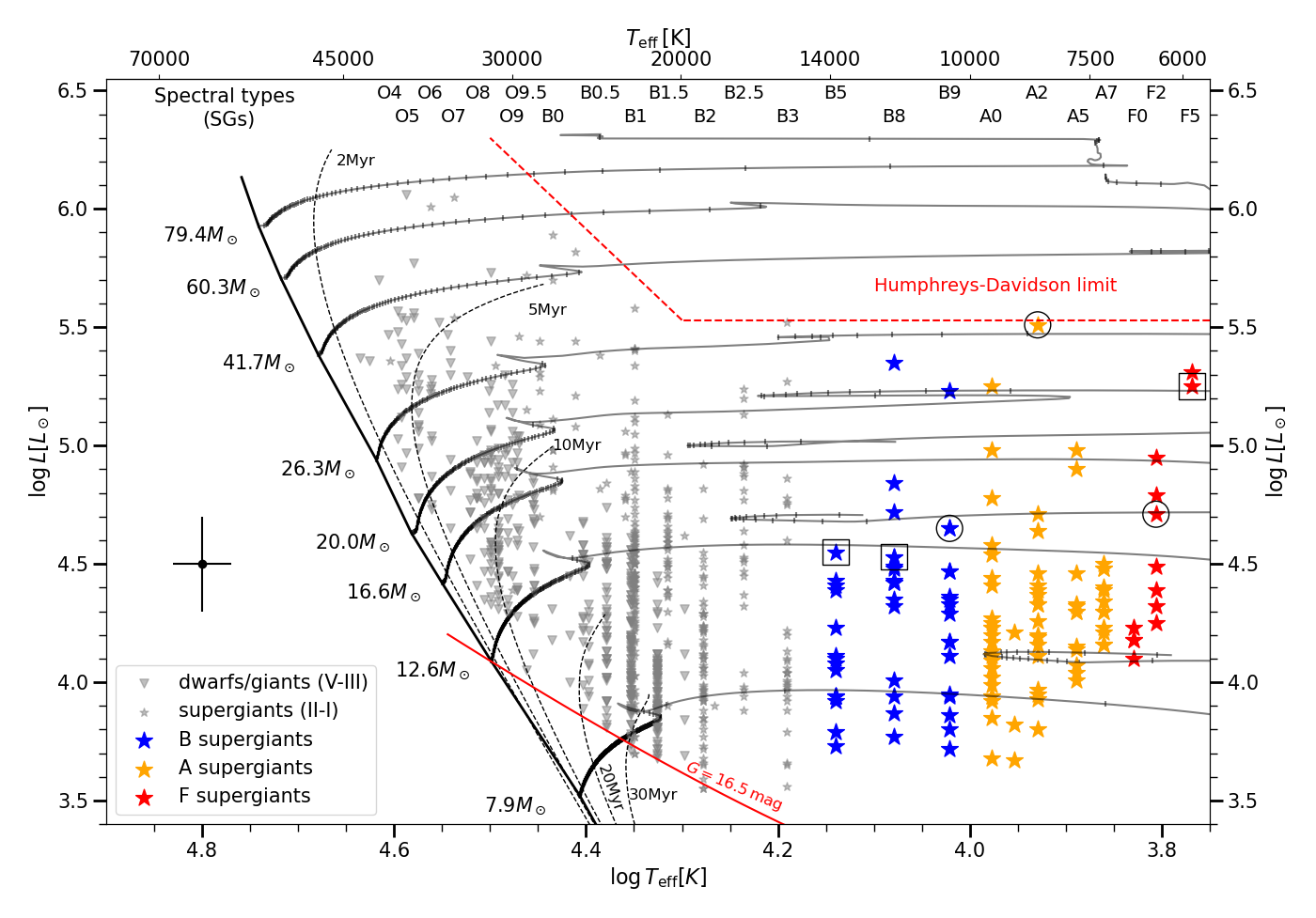}
\caption{Hertzprung--Russell diagram (HRD) of the BLOeM survey highlighting the BAF supergiant sample studied in this article with large star symbols colour-coded based on spectral type as indicated by the legend. 
The three targets with the black circles indicate the only targets in this sample that have a clear indication of binarity: BLOeM 4-072, BLOeM 6-006, BLOeM 6-008. 
The three targets with the black squares indicate targets that display binary motion characteristic of long orbital periods: BLOeM 1-114, BLOeM 3-106, BLOeM 8-001. 
Stellar parameters are determined using a calibration to spectral types by \citetalias{BLOeM-I}. 
Grey points show the entire BLOeM sample and stellar 
tracks from the extended grid of~\citet{Schootemeijer2019} with mass-dependent overshooting~\citep[described in App. B of][]{2021A&A...653A.144H}.
The black marker with error bars on the left gives an indication of the typical uncertainties on the stellar parameters.
The red solid line shows the magnitude limit imposed by the BLOeM survey target selection.
The red dashed line is the the Humphreys--Davidson (HD) limit, the cool part of which is shown as the revised down limit in the SMC from~\citet{Davies2018}.
}
\label{fig:hrd}
\end{figure*}

\begin{figure}
\centering
\includegraphics[width=0.5\textwidth]{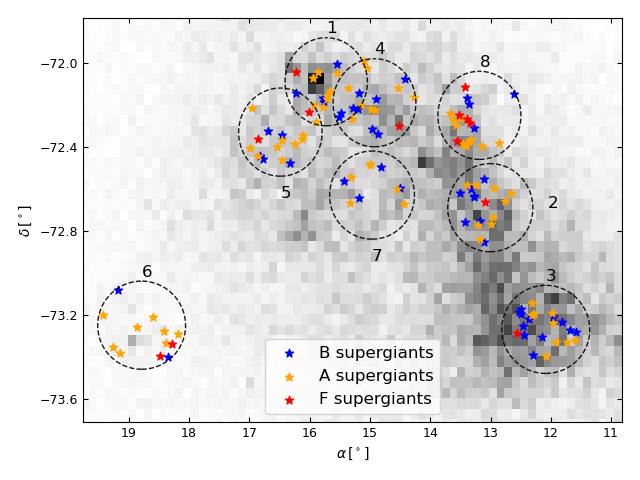}
\caption{Spatial location of the BAF-type supergiant targets in the SMC, overlaid on the density map of the Gaia~DR3 source catalogue ($G < 19$\,mag). Spectral types of the targets are colour coded as indicated. Black circles mark the locations of the 8 BLOeM fields.}
\label{fig:SMCDensity}
\end{figure}

The BLOeM campaign is implemented on
the Very Large Telescope (VLT) with the Fibre Large Array Multi Element Spectrograph~\citep[FLAMES;][]{Pasquini2002}. 
Briefly, it is used to observe $\sim$1000 massive stars distributed between eight FLAMES fields at different locations throughout the SMC using the LR02 grating, which covers the wavelength range 3960--4570\,\AA\ with a spectral resolving power ($R$) of 6200.
Figure~\ref{fig:SMCDensity} shows the spatial distribution of the BAF supergiants throughout the SMC on a density map from the Gaia DR3 source catalogue~\citep{2023A&A...674A...1G} while the number of sources per spectral subtype are shown in Table~\ref{tab:SpT}.
Within the eight BLOeM fields there are typically 16 BAF supergiants per field. Field 3 has the most at 26, followed by Field 8 with 22. 
As noted, while the full survey is delivering 25 epochs over 2\,yr, the currently available nine epochs were obtained over roughly 30--65\,d depending on field number.\footnote{The exact baseline per field is detailed in the appendix of~\paperI.} 

Data reduction and processing is described in \paperI, we note here that we use the cosmic ray cleaned, merged and normalized spectra in the present analysis.
While most sources have spectra for all nine epochs, a few sources have fewer spectra, as noted in Table \ref{tab:summary}. The median signal-to-noise ratio (s/n) of the individual spectra is 70, with an inter-quartile range of 40.
In addition to the quality control checks performed on the sample from \paperI, we performed some further comparisons with well known stellar catalogues including Gaia~Cepheid catalogues~\citep{2019A&A...622A..60C, 2023A&A...674A..17R}, 
finding that BLOeM~5-005 is 
the classical Cepheid SV*~HV~1954.
Cepheid variables display large intrinsic variability, confirmed here in our RV measurements, therefore we exclude this star from our discussion of the sample properties.

Table\,\ref{tab:SpT} and Figure\,\ref{fig:hrd} demonstrates that the sample is dominated by later B-type and A-type supergiants: the B8-A7 stars comprise $\sim$75\% of the sample and have a mass range of roughly 6--26\,M$_\odot$, the bulk of which have masses below $\sim$13\,M$_\odot$. In this estimate we have assumed that the source masses are those implied by the evolutionary tracks.
As expected, there are no sources above the HD limit, although a few sources are quite close to that boundary.

\begin{table}[]
    \centering
    \caption{Number of targets and median peak-to-peak RV measurements ($\Delta\varv_{\rm med}$) as a function of spectral subtype for the BAF supergiants.}
    \label{tab:SpT}
\begin{tabular}{c c c }
\hline
Spectral subtype & Number of targets & $\Delta\varv_{\rm med}$\\ 
& & [\kms]\\
\hline
F5~~\,  &  2       & 2.4\,$\pm$\,1.5 \\
F2~~\,  &  6       & 1.0\,$\pm$\,1.6 \\
F0~~\,  &  4       & 0.6\,$\pm$\,0.5 \\
A7~~\,  &  12      & 0.6\,$\pm$\,0.3 \\
A5~~\,  &  13      & 1.0\,$\pm$\,0.5 \\
A2~~\,  &  17      & 1.1\,$\pm$\,1.3 \\
A1~~\,  &  3       & 1.2\,$\pm$\,0.4 \\
A0~~\,  &  24      & 1.2\,$\pm$\,1.7 \\
B9~~\,  &  14      & 4.4\,$\pm$\,3.0 \\
B8~~\,  &  16      & 2.7\,$\pm$\,3.6 \\
B5~~\,  &  17      & 2.1\,$\pm$\,2.9 \\

\hline
\end{tabular}    
\end{table}

\begin{figure*}[htp]
\centering
\includegraphics[width=0.95\textwidth]{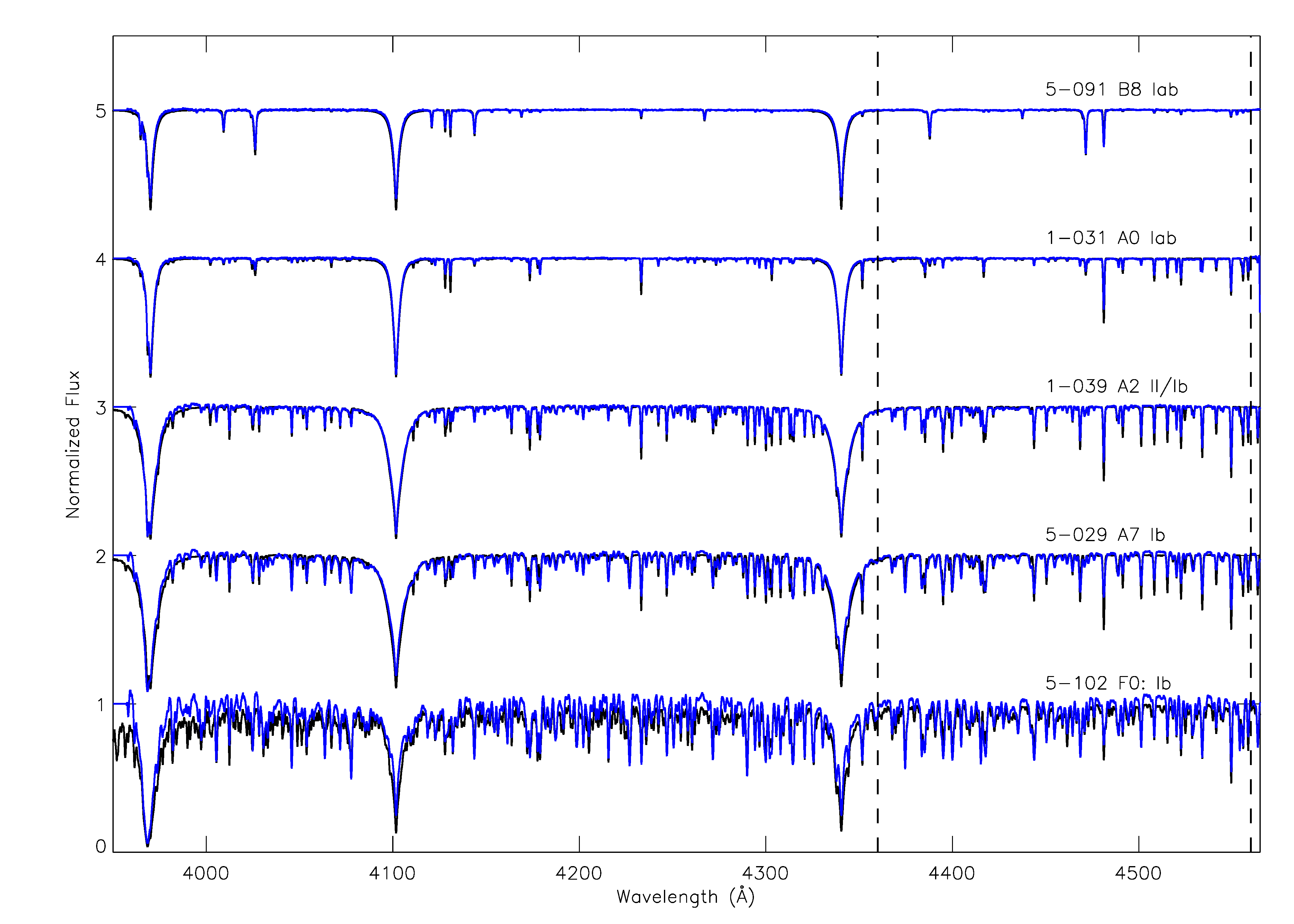}
\caption{Montage of a selection of BAF supergiant star spectra from BLOeM (in black), and best matching theoretical template spectra (blue) from CMFGEN/ATLAS model grids (see text). BLOeM IDs and spectral types are noted above each spectrum. Template spectra were used to determine an absolute RV scale, and one can see good agreement between line positions and strengths of the templates with the data. The vertical dashed lines indicate the spectral region that was used to derive the cross-correlation functions (CCFs) for all sources. Note the slight mismatch of the continuum for the F supergiant, BLOeM~1-102. For cool stars the automatic continuum placement is more difficult due to the line density of the spectra, however the CCFs are not impacted by this offset as we use normalized CCFs such that the mean signal is subtracted from each spectrum before cross-correlation. Note that the A--F supergiant line spectra are dominated by \ion{Fe}{II} and \ion{Fe}{I} lines \citep[for line lists see][]{venn1995,Venn1999}.}
\label{fig:Montage4}
\end{figure*}

\section{Methodology}\label{sec:methods}

Detecting binary motion for BAF supergiant stars presents a more challenging task than for earlier spectral types in that the expected periods are long, and hence the RV variations are small. 
As an indication, the minimum period possible for a mass ratio ($q$) of $q=0.5$ for the range of parameters indicated in the HRD are between roughly 8 and 350\,d, or 80 and 30\,\kms in semi-amplitude velocity (for circular orbits). 
Adopting $q=0.1$ changes the period range to 6--260\,d and the semi-amplitude velocity range to 20--8\,\kms\ (see Section~\ref{sub:configs} for further discussion).
As these observations span 30--65\,d, and that long-period binary systems are likely eccentric, 
we expect that the majority of the detected binaries to exhibit rather small velocity variations. 
Accordingly, in our RV analysis we pay particular attention to precision and potential systematic errors.


We adopt the cross-correlation approach to measure RVs, which is well suited to stars with many lines as is the case for the bulk of our sample. The precision improves as $\sqrt N$, where $N$ is the number of detectable lines.
Figure~\ref{fig:Montage4} illustrates the significant change in morphology and line density across the spectral types considered here.
We find that for some stars there is a small ($\sim$few \kms) instrumental shift between different spectral regions.
Therefore, we restrict the cross-correlation to 4360--4560\,\AA\ (indicated by two dashed lines in Figure~\ref{fig:Montage4}) to ensure we use a common wavelength region for all stars.
This mitigates against possible systematic errors in the wavelength solution but is sufficiently wide to include enough strong lines in the earliest spectra types.
Measurement of the RVs consists of two distinct parts: (i) determination of the relative RVs between separated observations; and (ii) determination of the zero-point of the RV scale for each star.

Zero-points are determined using low metallicity (Z) grids of CMFGEN models for B5--A supergiants covering $7250-14500$\,K in effective temperature, a log\,g range of 1.0~--~2.5, and 1/5 and 1/10 solar metallicity (Garcia et al, private communication). 
This was supplemented with a low-Z (1/5 solar) ATLAS9 grid for the F-stars \citep{howarth2011} as important opacity sources for the cooler objects are currently not implemented in the CMFGEN grid.
Important caveats include that the combined grid of models has a fixed Z, and also a fixed microturbulence value of either 5 or 10 \kms.
Hence, these grids do not provide tailored fits, but are sufficient for RV determination.
We find that RV zero-points depend slightly on the choice of model template, because of this we have cross-correlated each spectrum against all the models to find the best-matching template.
We emphasize that the objective here is not to find definitive stellar parameters, given the caveats above, rather it is to find the model in our pre-defined grid that best matches the line spectrum of each star. 
Nevertheless the effective temperatures derived from the best matching template are in good agreement with those derived from spectral type calibrations from \paperI, and are listed in Table \ref{tab:summary}.
For consistency we also list new values of the stellar luminosity assuming a mean extinction for the SMC from \citet{gordon2024}.
Inspection of Figure~\ref{fig:Montage4}, which shows some typical template matches, confirms that the cross-correlation regions are well modelled by the CMFGEN/ATLAS9 spectra.

Relative RV shifts between the nine epochs are determined by cross-correlating the normalized spectra against each other.
We then form an observational template spectrum by co-adding the separate exposures, taking care to use appropriate weights according to their respective s/n, and this is then cross-correlated against the model template to determine the RV zero point offset for the data set.
Cross-correlation uncertainties are calculated following \citet{zucker2003}, with some minor modifications. Further details of this and the cross-correlation process are provided in Appendix \ref{sec:errors}. 

The derived RVs and their uncertainties are listed in Table \ref{tab:rvels}, which shows that measured uncertainties can be as low as $\sim$0.1\,\kms with a median value of 0.5\,\kms. 
Consequently, it is important to have a clear understanding of the likely systematic errors that might be present in the data. 
Plotting measured RVs a function of time for stars grouped by their field number reveals that correlations are present, which are an indication that small common velocity or pixel offset exist between epochs for each field (see Figure~\ref{fig:perfield}).
As discussed in more detail in Appendix~\ref{sec:rv_corr}, we define a subset of sources for each field that are most strongly correlated, and by implication have RV variations below the limit of the correlated errors. 
By assuming these sources have a constant RV we compute median RV deviations at each epoch across all sources in each field. These corrections are applied iteratively to the measured data, with a few iterations typically sufficient to achieve convergence.
These final RV corrections are listed in Table~\ref{tab:corrections} and are applied to the measured values for all sources in the present analysis (including those reported in Table~\ref{tab:rvels}) and the subsequent discussion.
Note that a corollary to Table~\ref{tab:rvels} is that there is an as yet undetermined RV zero-point systematic uncertainty to the absolute RVs of order 1--2\,\kms.

Further insight into the nature of the RV variability of the sample can be obtained by computing some simple statistics for each source with results obtained from the corrected values summarized in Table \ref{tab:summary}.
In the subsequent sections we examine the mean RV and its standard deviation, the peak-to-peak RV variability ($\Delta\varv$, which is the difference between the maximum and minimum RVs) and the median error of the individual RV measurement errors. 
In addition, for each target we determine the significance of each pair of measurements using the equation,

\begin{equation} \label{eq:sig}
\sigma_{i,j} = \frac{|\varv_i-\varv_j|}{\sqrt{\sigma_i^2+\sigma_j^2}},
\end{equation} 

\noindent where $\varv_{i,j}$ and $\sigma_{i,j}$ are the RVs and their associated uncertainties at epochs
$i$ or $j$.
As discussed by \citet{Sana2013,Dunstall2015,Patrick2019} this quantity is often used in conjunction with a threshold condition, 

\begin{equation} \label{eq:rv_lim}
|\varv_i - \varv_j| > \Delta \varv_{\rm lim},
\end{equation}

\noindent where $\Delta \varv_{\rm lim}$ is chosen to reject possible intrinsic RV variations that are not caused by orbital motion but, for example, are the result of stellar activity such as pulsation.  

\section{Results} \label{sec:results}
\begin{figure}[htp]
\centering
\includegraphics[width=\linewidth]{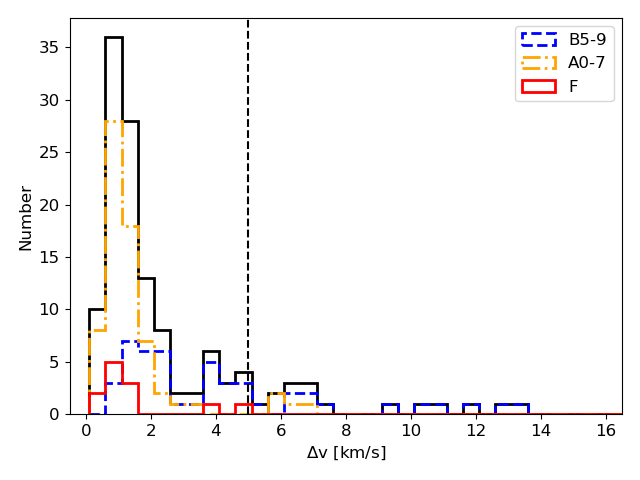}
\caption{Histogram of peak-to-peak radial velocity (RV) measurements ($\Delta\varv$) for the BAF supergiant sample. Different colours highlight the subsamples as indicated in the panel. The black-dashed line is the chosen threshold for binarity in the multiplicity analysis.
}
\label{fig:BAFhist}
\end{figure}

A key discovery from the analyses of the observed spectra is the absence of large-scale $\Delta\varv$ values for the entire BAF supergiant stars within the BLOeM sample. 
Figure~\ref{fig:BAFhist} shows the distribution of $\Delta \varv$ values as a function of spectral type, and Table~\ref{tab:SpT} shows the median values as a function of spectral subtype. 


\subsection{Binary detection}\label{sub:detection}


Equation~\ref{eq:sig} together with an appropriate choice of the threshold indicated in equation~(\ref{eq:rv_lim}) are typically used as criteria for the detection of binary candidates.
For example, \citet{Sana2013} adopted $\Delta \varv_{\rm lim}=20$\,\kms\ and $\sigma_{i, j}=4$ for O-type stars in the 30~Doradus region of the LMC, while \citet{Dunstall2015} adopted 16\,\kms\ and $\sigma_{i, j}=4$ for B-type stars in the same region. The latter choice reflects a somewhat better precision on RV measurements and smaller expected contribution from intrinsic RV variability~\citep[e.g.][]{2020A&A...643A.116D}.
For O-hypergiants and WNh stars~\citet{2023MNRAS.521.4473C} selected a larger limit of 30\,\kms, while maintaining the $\sigma_{i, j}=4$.

The choice of $\sigma_{i, j}$ (c.f. equation~(\ref{eq:sig})) is motivated by the sample size -- to exclude false positives -- and the level of significance attached to the detection.
For the BAF supergiant sample small uncertainties on the $\varv$ measurements mean that the choice of the $\sigma_{i, j}$ threshold does not affect the results significantly. 
Therefore we select $\sigma_{i, j}=4$.
Consequently, there are 23 sources (excluding the Cepheid) that meet this criteria.

The choice of the $\Delta \varv_{\rm lim}$ threshold (c.f. equation~(\ref{eq:rv_lim})) is driven by the need to exclude intrinsic RV variability as a result of pulsations or other physical mechanisms. 
For example, intrinsic RV variability of O-type supergiant stars was found to be significant by ~\citet{1996ApJS..103..475F}, and the vast majority of studied B-type supergiants are observed to be intrinsically variable (see \citealt{Bowman2019, 2020A&A...643A.116D}). However, RV variability is rather modest for most RSGs~\citep{2007A&A...469..671J,Patrick2019,2021MNRAS.502.4890D}.
From an examination of Figure~\ref{fig:BAFhist}, adopting even the lowest of the RV thresholds from the aforementioned previous studies in the OB-star regime results in the detection of no binaries in the BAF supergiant sample.
In an attempt to define a physically meaningful threshold for binary detection, \citet{Sana2013VFTS} assessed the distribution of $\Delta \varv_{\rm lim}$ as a function of stars that meet the threshold (see Figure~\ref{fig:CvsRV}). 
These authors argued that their results displayed a kink at $\sim$20\,\kms, which they attributed physical meaning and subsequently selected this as the threshold for binarity.
This prompted other authors to look for similar features, which are typically not found (for examples in this series see Fig. 5 of~\citet{BLOeM-BV} or Fig. 5 of~\citet{BLOeM-BI}).
We analysed this distribution for the BAF supergiant stars (and B5-9 and AF stars individually) and did not find a clear indication for a transition between intrinsic and orbital variability.
In lieu of a physically motivated choice of $\Delta \varv_{\rm lim}$, we select $\Delta \varv_{\rm lim}=5$\,\kms\ based on the distribution of observed $\Delta \varv$ values from the BAF supergiants.
Figure~\ref{fig:BAFhist} demonstrates that a choice of $\Delta \varv_{\rm lim}=5$\,\kms\ is a reasonable value to mitigate intrinsic variability.
We acknowledge that this limit is unsatisfactory.
To address this further may require full binary solutions for the candidates detected here but this is clearly outside the scope of the current article.
Similarly, in tandem, photometric constraints on intrinsic variability could potentially be used on a star-by-star basis to determine a physically motivated threshold $\Delta \varv_{\rm lim}$. 

Based on these criteria (i.e. adopting \ $\Delta \varv_{\rm lim}=5$\,\kms and $\sigma_{i, j}=4$) only 13 of the sample, or 10\% of the total BAF supergiant sample, meet both RV variability criteria to be classified as candidate binaries.
These stars are marked with the `$var$' flag in Table~\ref{tab:summary}.
Splitting the sample by spectral type we find the B5-9 type stars have an observed binary fraction of 25\,$\pm$\,6 (12/47)\footnote{The uncertainties for the percentages in this paragraph are determined using binomial statistics where the standard error is $\sigma = \sqrt{p\times (1-p)}/n$}.
For the A-type supergiants we find 4\,$\pm$\,2\,\% (3/69), and 8\,$\pm$\,8\,\% (1/12) for the F-type supergiants.
The lone binary candidate F-type supergiant, BLOeM~6-006, is also a borderline case. 
Further dissecting the B-type supergiants reveals no significant evidence for a trend as a function of spectral subtype.
Combining the AF stars we find 5\,$\pm$\,2\,\% (4/81).

For the B5-9 stars, it is likely that the specified 5\,\kms\ threshold for binarity is significantly contaminated by intrinsic RV variability (see Section~\ref{sub:varb}).
The AF stars appear to be less affected by intrinsic variability, therefore in the remainder of this article we split the sample to a consider the B5-9 and the AF supergiant stars separately. 
If we assume a more stringent limit of 10\,\kms\ for both samples in attempt to minimise the effects of intrinsic RV variability we find an observed binary fraction of $10\pm4$\,\% and 0\,\% for the B5-9 and AF samples, respectively.

\begin{table}[]
    \centering
    \caption{Range of orbital parameters and their assumed distributions considered in the bias correction calculations.}
    \label{tab:orb_params}
\begin{tabular}{l c l}
\hline
Orbital Parameter & Range & Distribution\\ 
\hline
$\log$ P/$d$ [dex] & 0.75 -- 3.5 & $\pi=+0.1\pm0.2$\\
$q$ & 0.1 -- 1.0   & $\kappa=0.0\pm0.5$\\
$e$ & 0 -- 0.9     & $\eta=-0.5\pm0.2$\\
M$_1 [M_\odot]$ & 6 -- 20 & $\gamma=-2.35$\\
\hline
\end{tabular}    
\end{table}
As a result of the sampling of observed epochs, the baseline of observation and the uncertainties associated with the RV measurements, the observed binary fraction should be considered a lower limit on the intrinsic binary fraction.
We account for these observational biases by simulating a population of binary systems using the Monte Carlo method developed by \citet{sana2012, Sana2013VFTS}.
Table~\ref{tab:orb_params} displays the ranges and assumed distributions of the orbital parameters that are used to determine the bias-corrected multiplicity fraction. 
The orbital period distribution is assumed given the result of~\cite{BLOeM-O}. 
The flat $q$ distribution is based on results from~\citet{Shenar2022TMBM} and~\citet{2022MNRAS.513.5847P}.
The mass function of the sample was tested using the stellar parameters determined in~\citetalias{BLOeM-I} and found to be consistent with a power law of exponent $\gamma =-2.35$.
This approach results in a bias-corrected multiplicity fraction of $\sim$30\% for the BAF supergiant sample.
However, as noted above we consider that the results for the B5-9 subsample are impacted by intrinsic variability, specifically by the prevalence of the $\alpha$~Cygni variables --- see Section~\ref{sub:varb}. 
Because of this, we split the sample at A0 and redetermine the statistics by taking into account that the AF stars are significantly larger than their B5-9 counterparts by considering orbital periods in the range 1.25~$<\log P/d [\rm dex] <$~3.5 (see Section~\ref{sub:configs}). 
This results in a bias-corrected multiplicity fraction of 8$^{+9}_{-7}$\% for the AF supergiants, which is thus less affected by intrinsic RV variability.
In addition, for the B5-9 subsample we perform the bias correction simulations for the observational limit of 10\,\kms, which results in an intrinsic binary fraction of 18$^{+20}_{-16}$\% for the B5-9 supergiants.

The choice of the assumed orbital property distributions and their uncertainties clearly has an impact on the final quoted binary detection statistics. 
Our chosen distributions largely represent those appropriate for unevolved stars of similar masses and assume no binary evolution, for consistency with other studies in the BLOeM campaign.
If the population of BAF supergiants is dominated by binary interaction products, the bias-corrected multiplicity fraction is likely to be inaccurate.
To assess the impact of this we repeat the bias-correction simulations with a range of underlying distributions with for example a split $q$-distribution~\citep{Moe2017}, which favours more low-$q$ systems. 
This results in a bias-corrected multiplicity fraction of 8$^{+9}_{-8}$\% for the AF supergiants, which is in good agreement with the above results.
This gives us confidence that the final results are robust to the exact choices of the orbital parameter distributions.


\subsection{Constrains on intrinsic variability from visual inspection} \label{sub:varb}

The size and spectral coverage of the BAF supergiant sample combined with the BLOeM campaign design -- and the relative lack of RV variability found in the previous section -- provides an excellent opportunity to characterise the intrinsic variability of BAF supergiants in the SMC.
The distribution of $\Delta \varv$ values for the BAF supergiant sample as a function of spectral type, see Figure~\ref{fig:BAFhist}, clearly shows the scarcity of large $\Delta \varv$ values at all spectral types.
The $\Delta \varv$ measurements cluster around a value of 1\,\kms\ with a sharp drop beyond 2\,\kms, and perhaps a small over-density near 4-6\,\kms, mostly due to the B-type stars. 
In particular, for the AF supergiant stars there are no $\Delta \varv$ values larger than $\sim$6\,\kms.
Table~\ref{tab:SpT} shows the typical $\Delta\varv$ values as a function of spectral type.

\begin{figure*}[htp]
\centering
\includegraphics[width=0.95\linewidth]{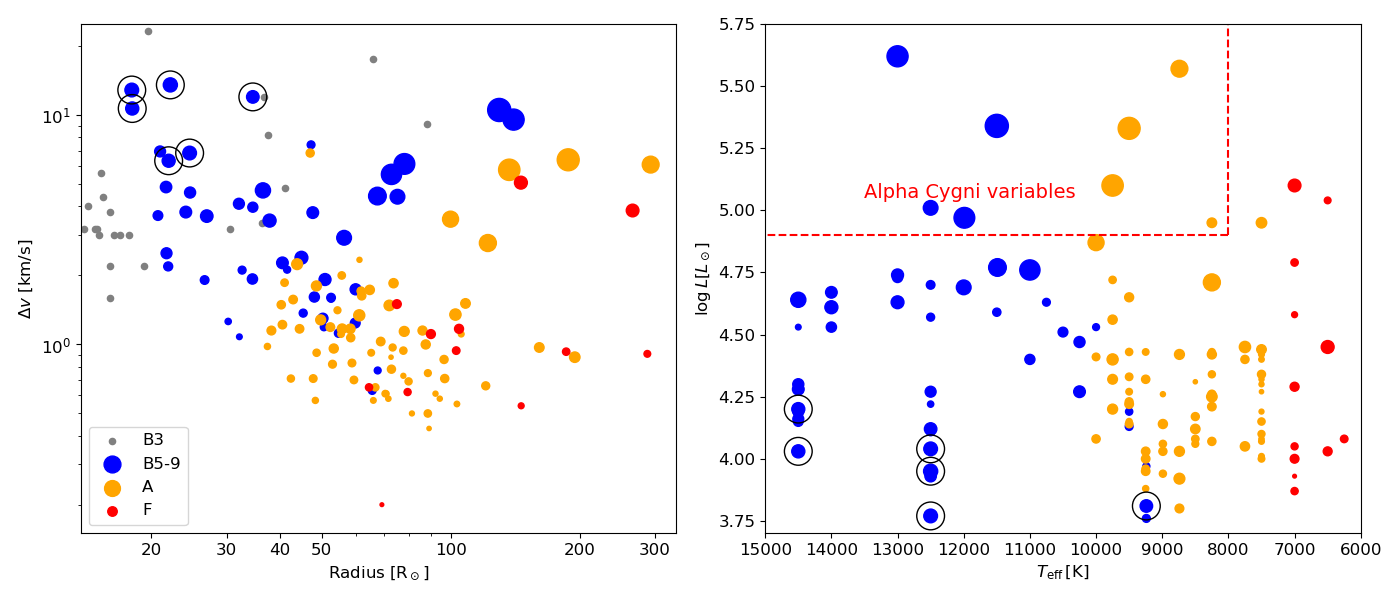}
\caption{Left panel: Peak-to-peak RV ($\Delta \varv$) as a function of stellar radius for the BAF supergiants and the B3 supergiants of~\citet{BLOeM-BI}. 
Radii are determined using effective temperatures and luminosities from Table\,\ref{tab:summary}. 
The size of the symbol in both panels are determined by $\Delta\varv/\sigma_{\Delta\varv}$.
Black circles mark stars the cohort of variable late-B-type bright giants.
Right panel: HRD for BAF supergiant sample using the stellar parameters listed in Table~\ref{tab:summary}, symbols have the same meaning as in the left hand panel. Note that the sources with largest $\Delta\varv/\sigma_{\Delta\varv}$ tend to be the most luminous and hottest stars. 
}
\label{fig:p2p_radius}
\end{figure*}

To investigate the intrinsic RV variability, Figure~\ref{fig:p2p_radius} shows $\Delta\varv$ as a function of stellar radius together with the B3 supergiants from~\citet{BLOeM-BI}.
The latter are likely to be post-main sequence stars as discussed~\citet{BLOeM-BI} and are expected to display similar RV properties to the B5-9 stars. 
The sizes of the points in Figure~\ref{fig:p2p_radius} are determined by the significance of the $\Delta\varv$ measurement, which is defined as $\Delta\varv/\sigma_{\Delta\varv}$.
This distribution reveals a clustering of points that suggest a trend of decreasing $\Delta\varv$ with increasing radius (and, by inference, with decreasing effective temperature).
This trend is driven primarily by the improving precision as one moves to cooler stars that have more spectral lines. Moreover, many of the B3 and B5 supergiant stars have lines broadened by rotation that leads to larger RV uncertainties at the hot end of the distribution.  
However, the stars with the largest $\Delta\varv/\sigma_{\Delta\varv}$ are discrepant from this general distribution. 



In the right-hand-panel of Figure~\ref{fig:p2p_radius}, we see that the stars with the most significant RV variability and largest $\Delta\varv$ values, are among the most luminous BAF supergiants.
\citet{kaufer1997} presented a detailed analysis of long term monitoring data for photospheric lines of six luminous supergiants in the B7--A2 spectral domain (i.e. $\alpha$ Cygni variables that include $\alpha$ Cyg and $\beta$ Ori) and found significant RV variability with $\Delta\varv$ values in the range 15 to 20\,\kms.
These authors tentatively attributed this variability to non-radial gravity-mode pulsations, though this is debated \citep[see for example][]{saio2013}. 

For the most luminous stars that occupy a similar region in the HRD to the sample discussed by \citet{kaufer1997}, 
it seems likely that we are detecting $\alpha$ Cygni variability due to pulsations rather than orbital motion. 
Here we assign seven stars an $\alpha$~Cygni flag in Table~\ref{tab:summary}, which is based on their high luminosity and significant RV variability.  
It is interesting that there is a lack of strongly variable, $\alpha$ Cygni-like, sources among the AF supergiants. 
This is perhaps indicative of their differing luminosity-to-mass ratios, and/or perhaps hinting at a different evolutionary path. 
While we assign only seven stars the $\alpha$~Cygni designation, it is likely that our selection misses a significant number of candidates at lower luminosities.

\begin{figure*}
\centering
\includegraphics[width=\linewidth]{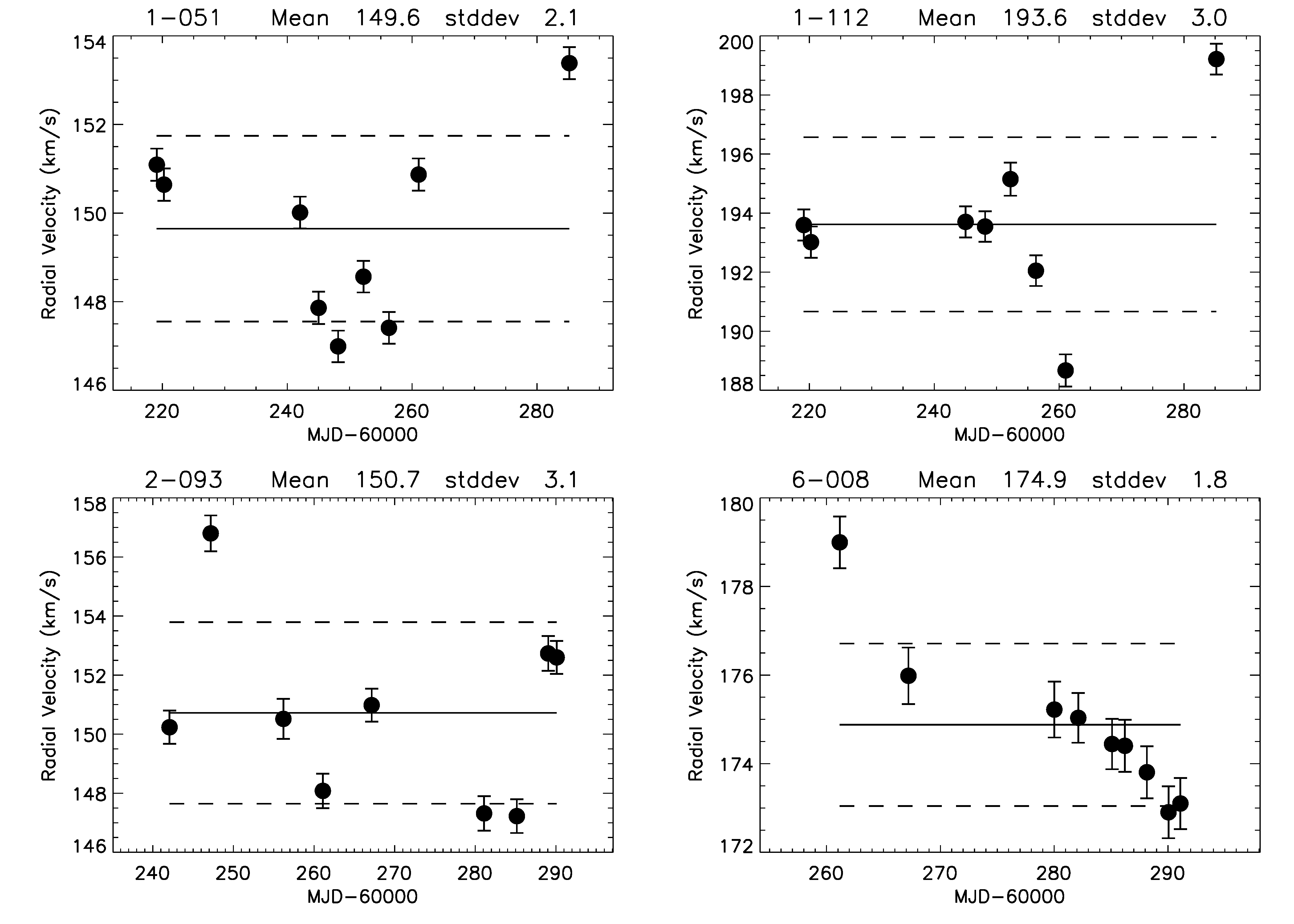}
\caption{Radial velocity (RV) time series for four of the targets that display RV variability above the chosen thresholds (see text), with spectral types of A\,Ia (BLOeM~1-051), B9\,Ia (1-112), B8\,Ia (2-093) and A2\,Ia (6-008). Identifications are listed in the panel headings, as are the mean RVs and their standard deviations, which are indicated in each panel by solid and dashed lines, respectively. Note that only 6-008 displays evidence of a long-term RV trend that might be a signature of binary motion. Due to their large radii, all of these stars have minimum periods of $\sim$100\,d or more.}
\label{fig:rv3panel}
\end{figure*}

To test whether the observed RV variability is a result of orbital motion or pulsation 
we visually check for long-term trends in the RV time series.
Figure~\ref{fig:rv3panel} illustrates three examples of RV variability that are clearly subject to variations on short timescales (days) that would be consistent with the findings of \citet{kaufer1997} for pulsations \footnote{Three additional sources with similar properties to these are are BLOeM~2-092, 4-072 and 8-010.}.
The fourth panel in that figure, for BLOeM~6-008, is an example of a potential binary showing a long-term trend with clearly smaller short timescale variability. 

These four sources have high luminosities and, therefore, large radii resulting in minimum periods for $q=0.5$ of 100\,d or more. 
Fitting orbital solutions to the RV time series data using the program RVFIT \citep{rvfit} yielded only spurious solutions for the first three sources. The solution for BLOeM~6-008 indicates a period of $\sim$180\,d, the fit is poorly constrained, which indicates that the full 25-epoch coverage of the BLOeM campaign is required to further our knowledge of this candidate binary system. 
In fact only two other stars in the $var$ category (c.f. Table~\ref{tab:summary}) exhibit RV variability that can be considered binary motion, namely BLOeM~4-072 (B9\,Ia) and 6-006 (F2\,:).
Time series RVs for all 14 sources flagged as $var$ in Table~\ref{tab:summary} are presented in Appendix\,\ref{sec:rv_plots}. Visual inspection of the full sample reveals only three additional sources that may be compatible with a long-term trend; BLOeM~1-114, 3-106 and 8-001. These objects are flagged as `LP' in table \ref{tab:summary}. 
Running RVFIT for all these systems resulted in similar poor results as for the $var$ sample.

In addition, based on further inspection of Figure~\ref{fig:p2p_radius}, we highlight a second cohort based on their similar $\Delta \varv$ values, effective temperatures and luminosities. This grouping consists of late B-type bright giants. The stars that we consider members of this cohort are BLOeM 3-092, 3-102, 4-004, 5-086, 6-007 and 6-097. These stars are highlighted with black circles in Figure~\ref{fig:p2p_radius}. Interestingly, all six objects appear to have significant rotational broadening.


In summary, it is clear that the observed and bias-corrected multiplicity fractions of BAF supergiant stars are seriously compromised by the intrinsic  variability of the most luminous systems. The true binary fraction is significantly smaller.

\begin{figure*}
\centering
\includegraphics[width=\linewidth]{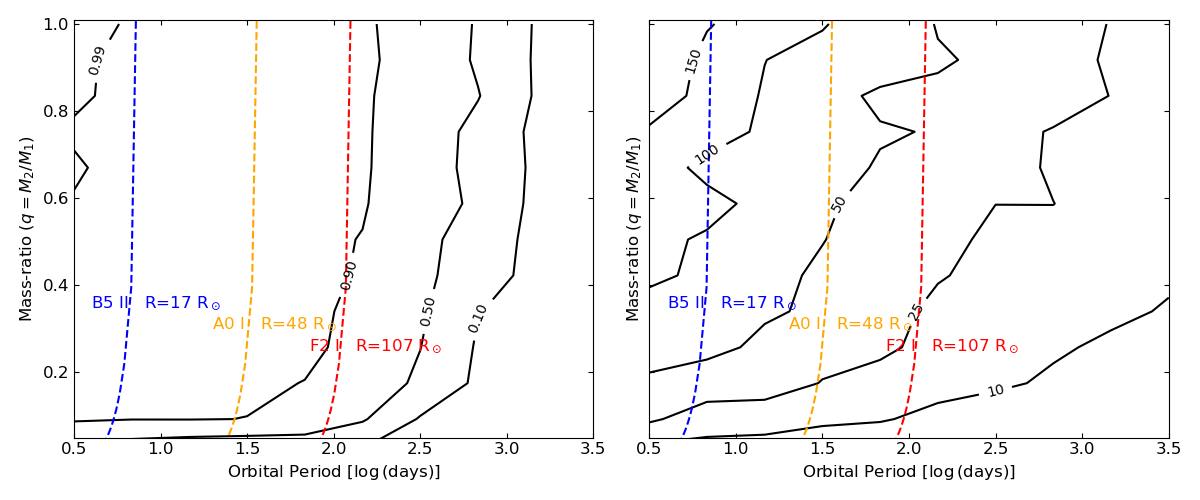}
\caption{\textbf{Left:} Orbital period shown against mass ratio ($q$) for simulated binary systems in the orbital period ranges that are used to correct for our observational biases. 
Solid black contour lines show the probability of detection at the 99, 90, 50 and 10\,\% level given the temporal baseline and typical uncertainties of the BLOeM data where each grid point is the median of 10\,000 simulated systems. The detection criteria used in these simulations are matched to those of the observations (i.e. $\Delta \varv$=5 \kms\ and $\sigma=4$). The coloured dashed lines show the minimum allowed orbital period for three representative examples as a function of the mass-ratio of the system.
\textbf{Right:} The same simulations where the solid contour lines now show lines of constant semi-amplitude velocity on the 150, 100, 50, 25 and 10\,\kms\ level.}
\label{fig:detection}
\end{figure*}

\subsection{Constraints on detectable orbital configurations} \label{sub:configs}
The result of the low detected binary fraction for the BAF supergiant stars naturally brings the question of how sensitive is the BLOeM survey to detecting binary systems in the appropriate orbital period range. 
To assess this, we simulate populations of binary systems using the parameters given in Table~\ref{tab:orb_params}.
In these simulations each system is assigned a random inclination angle with respect to the observer and we generate RVs with a baseline, sampling and uncertainties typical of the BLOeM observations. 
We then test whether or not each system would meet the criteria to be classed as a binary, similar to how we applied our multiplicity analysis described in Section~\ref{sub:detection} (i.e. $\Delta \varv$=5 \kms\ and $\sigma=4$).
Figure~\ref{fig:detection} shows two findings from this analysis for the P-$q$ parameter space, in which the solid contour lines in the left-hand panel indicate the probability of detection from 10\,000 binary simulations per grid point.
Whereas, in the right-hand panel the solid lines show constant semi-amplitude velocity values from the simulated systems.
Also shown is the minimum period possible for three stars, BLOeM 3-009 (B5\,II), 4-075 (A0\,I) and 8-089 (F2\,:) for a range of $q$ values. 
Stellar parameters are taken from Table~\ref{tab:summary} and estimates of stellar masses from the evolutionary tracks. 
Note that these three stars are clustered around the median of the luminosity distribution 
(with $\log(L/L_{\odot}$ in the range 4.3--4.5) and radii typical of their spectral types. Hence the constraints illustrated in Figure~\ref{fig:detection} and discussed below are even more stringent for the more luminous stars.

In the left-hand panel of Figure~\ref{fig:detection} it is apparent that the observations are sensitive to orbital periods up to hundreds of days, but sensitivity falls off significantly beyond 1000\,d. 
For example for the B5 stars, binary systems would be detected in the $5 < P < 100$\,d range at above the 90\% confidence level for the full range of $q>0.1$. 
In contrast, for the 12 F-type supergiant stars, the parameter space of orbital configurations that reach the 90\,\% level of detection probability is very small and as such, our true detection probability for the F-type stars is significantly smaller than that of the B-type stars.
As the assumed $\log$\,P and $q$ distributions are roughly flat in nature (for $q>0.1$), the area to the right of the dashed lines in Figure~\ref{fig:detection} illustrates the true detection probability for each target.
As a representative example, the observations are around twice as likely to be able to detect a binary motion for a B5 star than an A0.


Taken together, Figure~\ref{fig:detection} clearly shows that binaries with orbital periods within the $5 < P < 100$\,day range and having $q>0.1$ should produce a strong signal within the current observational setup, but are not observed.

Beyond an orbital period of around 300\,d a significant fraction of companions go undetected based on orbital motion.
While the bias-correction simulations account for such systems in the statistics determined in Section~\ref{sub:detection}, it is important to describe the types of systems that are undetected in the current observational setup. 
The UVIT survey allows us to fill in such systems at $<\sim$5\,M$_\odot$ (see Section~\ref{sub:uv}) but the BAF sample could still harbour undetected long-period companions.
This includes main-sequence companions as well as binary interaction products such as stripped stars and compact companion systems.

\subsection{Binary systems detected via UV excess} \label{sub:uv}

\begin{figure}
    \centering
    \includegraphics[width=0.9\linewidth]{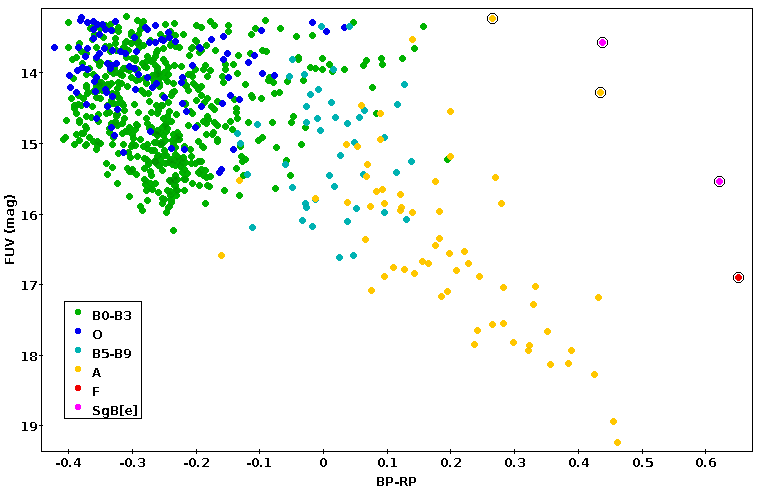}
    \caption{Colour-magnitude diagram using the 754 BLOeM sources found in the UVIT FUV catalogue \citep{uvit2024}, colour-coded by spectral type. The FUV magnitude was obtained with the F172M filter, while the $BP-RP$ color is from $Gaia$ $B_{p}$ and $R_{p}$ filters. The five outliers described in Section~\ref{sub:uv} are flagged with black circles. }
    \label{fig:uvit}
\end{figure}

\citet{uvit2024} published a far ultraviolet (FUV) catalogue (Hota, private communication) for the SMC using data from the Ultra Violet Imaging Telescope (UVIT) onboard of $AstroSat$ and all but a handful of BLOeM targets overlap with its footprint. 
Cross matching with that catalogue using a 0.5\arcsec\ matching radius we find 754 BLOeM sources with FUV magnitudes.
Figure\,\ref{fig:uvit} displays the matches in an optical-FUV colour-magnitude diagram (specifically $B{_p}-R{_p}$ versus FUV).
Only three BAF sources are outliers in this diagram (BLOeM~5-109 (A7\,Iab), BLOeM~6-008 (A2\,Ia) and BLOeM~4-006 (F2:).  
We assume that these sources with UV excesses are the result of a hot companion.
The FUV magnitudes of these sources are consistent with main-sequence companions in the mass range 15--20\,$M_{\odot}$ \citep[see Figure 5 of][]{2022MNRAS.513.5847P}, which may make the BAF supergiant the less massive component of possible binary systems.
However, the detection limit of the \citet{uvit2024} catalogue corresponds to a main sequence mass of around 5\,$M_{\odot}$, which may explain the low overall detection rate.
Note that BLOeM~6-008 is also a source that exhibits a long-term trend in its RV time series, as shown in Figure\,\ref{fig:rv3panel}.
BLOeM~{4-006} was denoted as a possible SB2 system, denoted as `SB2?' in \paperI, however its RV time series is effectively constant within the errors, hence this may be either a chance alignment or a very long-period system. BLOeM~5-109 also exhibits an RV time series that is constant within its errors. 
In addition to these three sources, the two B[e] supergiant stars that are in the BLOeM sample are also outliers, also shown in Figure~\ref{fig:uvit}.

A cross-match with the yellow supergiants (YSGs) in
\citet{2024arXiv240617177O}, who searched for UV excesses from Swift data, resulted in four stars in common: BLOeM~6-009, 6-002, 6-052, and 5-109. 
Of these, both BLOeM~6-009 and 6-052 are found to have UVIT FUV magnitudes commensurate with their spectral types. Whereas BLOeM~6-002 is not recovered, despite it being within the UVIT footprint.
Given the smaller point spread function (PSF) of UVIT compared to Swift -- a full-width half-maximum of 1\arcsec versus 2.5\arcsec -- it is possible the discrepancy is a consequence of the larger PSF of Swift detecting a nearby star. 
We note that BLOeM~{5-109} is the only source detected in the \textit{uvm2} filter from the Swift data, 
accordingly we only flag the UVIT UV excess sources in Table~\ref{tab:summary}.

\section{Discussion}
\label{sec:discussion}
\subsection{Direct evolution from main-sequence}
A general conclusion of our analysis (c.f. Section~\ref{sub:configs}) is that there appears to be a `period gap' up to $\sim$10$^{2}$\,d in the BAF supergiant sample. Specifically, there is a dearth of `short' period systems with mass ratios above 0.1. 
An important implication of this finding is that it provides constrains on the evolution of the system evolving red-wards across the HRD.  
To address this issue we compare our sample with that of
other BLOeM studies: \citet{BLOeM-BI} finds a bias-corrected multiplicity fraction of 40\,$\pm$\,4\,\% for the B0-3 supergiants, while \citet{BLOeM-BV} finds $80\pm8$\,\% for the B0-2 giants and dwarfs, and \citet{BLOeM-O} finds $70^{+11}_{-6}$\%\ for the O-type stars.

\begin{figure*}[htp]
\centering
\includegraphics[width=0.45\linewidth]{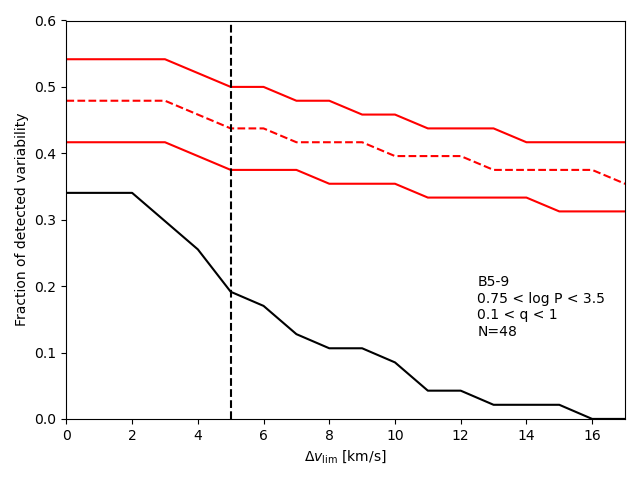}
\includegraphics[width=0.45\linewidth]{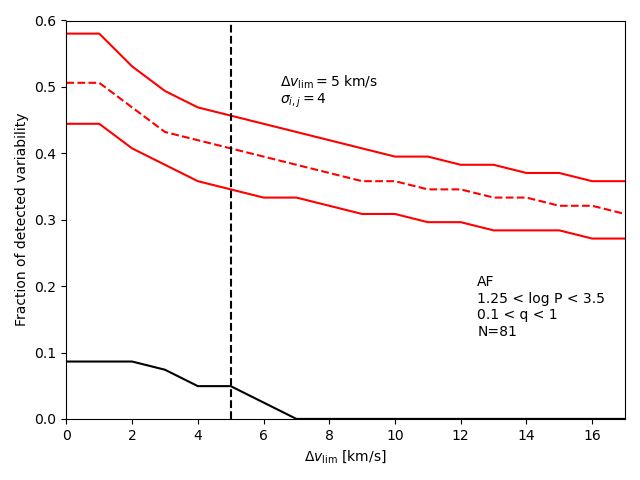}
\caption{Fraction of detected binarity as function of $\Delta \varv_{\rm lim}$. The black solid line in the left panel shows the B5-9 supergiants and the right panel is the same but for AF supergiants.  
The red dashed lines show the simulated binary fraction assuming an intrinsic binary fraction on the main sequence of 75\,$\pm$\,10\,\% -- to roughly match that of~\citet{BLOeM-BV} -- which is evolved to remove binary systems that would have since interacted based on the current sizes of the stars.
The solid red lines show the upper and lower bounds of the assumed uncertainty on the intrinsic multiplicity fraction.}
\label{fig:CvsRV}
\end{figure*}

To test the agreement between the multiplicity fraction of the different BLOeM samples we simulate populations of binary systems with ranges of orbital configurations and underlying distributions following Table~\ref{tab:orb_params} to account for the differences in the allowed orbital configurations for the BAF supergiant sample, which we split into two bins: the B5-9 supergiants and the AF supergiants.
We created simulated observations to match the sample size, baseline and uncertainties of each of the two subsamples. 
Figure~\ref{fig:CvsRV} shows a comparison between the observed results (black solid lines) and simulations of populations of binary systems whose bias-corrected multiplicity statistics are determined on the main sequence by~\citet{BLOeM-BV} and evolved to match the different samples (red dashed lines).
One would expect that the black and the red dashed lines to overlap significantly, particularly at large $\Delta \varv_{\rm lim}$, if the simulations agreed with the observations.
However, these comparisons demonstrate that the simulated population of binary systems evolved from the main-sequence drastically over-predicts the level of high-amplitude RV variability caused by binarity.
To illustrate this further we apply the same binary threshold criteria to the simulated data, $\Delta\varv_{\rm lim} = 5$\,\kms and $\sigma_{i, j}=4$, to predict an observed binary fraction for the AF supergiant stars of 40$^{+5}_{-6}$\,\%.
This is clearly significantly larger than the observed binary fraction of 5\,$\pm$\,2\%.
This quantitatively demonstrates that the origin of the BAF supergiants cannot be explained by a direct evolution from the main-sequence. 
Here we restrict the discussion to the AF supergiants because of the uncertain contribution from intrinsic RV variability for the B5-9 sample, although one can see from the left-hand panel of Figure~\ref{fig:CvsRV} that even without accounting for intrinsic RV variability, the simulations over-predict the observed multiplicity fraction.
Altering the simulations by experimenting with different values for the underlying $q$-distribution to favour more low-$q$ systems changes the appearance of the red lines in Figure~\ref{fig:CvsRV}, but does not alter the key result that the simulations cannot explain the absence of large-amplitude orbital motion.

As noted in the introduction, \citet{burki1983} find a binary fraction for F-type supergiants of $\sim$18\% in the Milky Way. However, their observing window was 5 years, such that all but a few of those binaries have periods in excess of this time span, and all have RV variations less that 5--10\,\kms\ (see their Figure 2a and Table 7). 
Hence, given our current observational baseline, it is unlikely that we would have detected any of their sources as RV variable. 
We conclude that our results are consistent with \citet{burki1983} given the differences in baseline and, hence, orbital period range.
An important corollary to that work is that such long periods, in excess of 5 years, and small RV amplitudes imply a very low binary mass function, probably indicating solar or sub-solar companions.

\subsection{Comparison with RSGs}
Assuming single star evolution, our sample are either evolving towards the RSG stage, or have previously been RSGs
\citet{2022MNRAS.513.5847P} studied the multiplicity statistics of RSGs ($\sim$4500~--~3800\,K) in the SMC with UV photometry and found a multiplicity fraction 18\,$\pm$\,4\,\% for orbital periods approximately in the range $3 < \log P [{\rm day}] < 8$, and $q$ in the range with 0.3\,$< q <$\,1.0.
They also argued, based on a comparison with multiplicity results from B-type stars in the LMC~\citep{Dunstall2015}, that the orbital period distribution drops drastically around $\log (P/{\rm day}) > 3.5$. 
The $q$-distribution that was assumed by~\citet{2022MNRAS.513.5847P} is the same as used here (i.e.\ flat), which means that we can account for the differences in the $q$ range considered.
Taking this into account results in a bias-corrected multiplicity fraction of the AF supergiant stars of 9\,\%.
This is smaller than the multiplicity fraction for the RSG population, but compatible at the 3-$\sigma$ level.
Here, as in the previous subsection, we restrict the discussion to the AF supergiant stars because of the uncertain contribution from variability for the B5-9 sample.
If the inference that the orbital distribution falls for periods longer than $\log P \sim 3.5$ is correct, this can be interpreted as evidence that the multiplicity properties of the AF supergiant stars are inconsistent with being pre-RSG objects as their binary statistics are incompatible with undergoing blue loops in the HRD. 
The lack of BAF supergiant sources with a FUV excess would also be consistent with this scenario. 


\subsection{Binary interaction products}
Having considered mainly single stellar evolutionary pathways for the origin of the BAF supergiant sample so far, we turn our attention to potential binary pathways.
For example, sources like the `puffed up' stripped star plus OB-star system VFTS\,291 \citep[][spectral type B5\,Ib-II]{villasenor2023}, which has an 108\,d orbital period and semi-amplitude RV of $\sim$94\,\kms, are clearly not present in the BAF supergiant sample. 
In addition, while the current spectral type of the stripped star `primaries' of systems such as LB-1 and HR\,6819 \citep[][periods of $\sim$79\,d and $\sim$40\,d]{Shenar2020LB1,Bodensteiner2020} are early B-type stars, their precursors would likely have been in the effective temperature range of BAF stars.
Given the expected low mass of such `primaries' (i.e. the optically brightest component), and relatively high mass of the secondaries (i.e. $q>1.0$), one might expect that the detection probability of such systems would be high, as implied by Figure\,\ref{fig:detection}. 
On the other hand, the expected number of such systems depends on the expected lifetime of this phase.
\citet{dutta2024} discussed the dependence on mass, luminosity and metallicity on such short-lived post-interaction evolutionary phases. From binary population synthesis they argued that as many as 0.5--0.7\% of systems with $\log L/L_{\odot}>3.7$ may be puffed up stars.  
One might therefore expect only a handful of such systems in the $\sim 1000$ BLOeM sources, the bulk of these among the more numerous O and early-B stars. 

Another important product of binary evolution are OB+BH binaries which are expected to comprise as much as $\sim$5\% of all OB stars in binaries \citep{langer2020}.
A small number of such systems have been identified observationally~\citep{2022A&A...664A.159M,2022NatAs...6.1085S}.
The LMC theoretical period distribution from~\citet{langer2020} has a significant fraction of systems with long periods, peaking at just over 100\,d, and with semi-amplitude velocities  of as much as 50\,\kms.
The lack of detection of such long-period compact companions is supporting evidence that the BAF supergiant sample cannot be explained by stars that have directly evolved from the MS. 

Furthermore, the SMC hosts a large number of high-mass X-ray binaries, (HMXBs) the bulk of these, $\sim$70, being Be X-ray binaries (BeXRB) comprising late O- to early B-type primaries and orbital periods of a few tens to a few hundreds of days \citep{mcbride2008, townsend2011, coekirk2015, haberl2016}. Only one system, SMC\,X-1, has a blue supergiant donor star, Sk\,160 (O9.7Ia+). The compact object in SMC\,X-1 is a neutron star. 
The range of primary spectral types among HMXBs in the SMC and mass functions implies $q$ values in the range $\sim$0.05--0.30, which is within our sensitivity range. 
To illustrate this, Figure~\ref{fig:detection} displays detection probabilities for orbital configurations down to $q=0.05$. 
This demonstrates that low-$q$ systems remain detectable at the 90\,\% detection level with orbital periods in the range 5 < P < 30\,d. 
Therefore, we might expect to detect long-period BeXRB (or OB plus neutron star binaries) that have not yet undergone a common envelope phase. We note that synthesis of the BeXRB binary population in the SMC also contains a tail of longer periods than are detected currently \citep{vinciguerra2020}. 
One might conclude therefore that the BeXRB population does not survive into this part of the HRD.
Finally, the scarcity of binaries in the BAF supergiant population is consistent with predictions of stellar mergers, occurring either on the main-sequence or in the RSG regime \citep{Justham2014, 2017MNRAS.469.4649M,schneider2024}, which predict single star products as blue supergiants in the BAF effective temperature range\citep{menon2024}. 



\section{Summary and conclusions}
\label{sec:conclusions}
In this article we have investigated the intrinsic RV variability and multiplicity properties of the supergiant stars ranging from spectral types B5 to F5 in the Small Magellanic Cloud (SMC).
We find a dearth of short-period binary systems ($5 < P < 100$\,d) for the BAF supergiants in the BLOeM campaign.
Evidence for this is the significantly smaller observed and bias-corrected binary fractions for the BAF sample compared to that of main-sequence stars \citet{BLOeM-O,BLOeM-BV}, and the small peak-to-peak RV values ($\Delta\varv$) observed.
This is in contrast to the study of the B0-3 supergiant stars in the BLOeM campaign, which show large $\Delta\varv$, which is indicative of short-period binary systems.
We find an observed multiplicity fraction of between 10 and 25\,\% for the B5-9 supergiants, which is likely affected by intrinsic variability, such as pulsations (e.g. $\alpha$ Cygni variables).
We find an observed multiplicity fraction of 5\,$\pm$\,2\,\% for the AF supergiant stars. 
By taking into account observational biases we determine the intrinsic multiplicity fractions are less than 18\,\% for the B5-9 supergiants and 8$^{+9}_{-7}$\% for the AF supergiants.
We caution, however, that these results, particularly for the B5-9 sample, are likely significantly affected by uncorrected intrinsic RV variability, which if accounted for would result in smaller intrinsic multiplicity fractions.

  
Our key conclusions are as follows:
\begin{itemize}
    \item[i.] We examine the evolutionary status of the BAF supergiants via a comparison with main-sequence stars in the BLOeM campaign from~\citet{BLOeM-BV} and conclude that the BAF supergiant stars are inconsistent with a direct, red-ward evolution from the main-sequence. This suggests the BAF supergiants were either born as effectively single stars or are binary interaction products. We compare our results with multiplicity statistics for red supergiants (RSGs) in the SMC and conclude that the AF supergiants are inconsistent with being pre- or post-RSG objects. While the origin of both groups of supergiants remains unclear, our results are able to rule out multiple evolutionary pathways. 
    \item[ii.] By assessing the BLOeM spectra and RV time series of the detected binary systems we conclude that, remarkably, only a small fraction of stars display convincing orbital solutions. 
    In addition, there are no systems where a definite orbital period can be determined.
    This suggests that the true multiplicity fraction of both the B5-9 and the AF supergiants is lower than 15\,\%, but inconsistent with zero, based on UV detected companions and indications of long orbital periods for a handful of systems.
\end{itemize}




\begin{acknowledgements}
We thank Sipra Hota for kindly sharing the SMC UVIT catalogue prior to publication.
LRP, FN. and FT acknowledge support by grants
PID2019-105552RB-C41 and PID2022-137779OB-C41 funded by
MCIN/AEI/10.13039/501100011033 by "ERDF A way of making
Europe". 
LRP acknowledges support from grant PID2022-140483NB-C22 funded by MCIN/AEI/10.13039/501100011033.
TS acknowledges support by the Israel Science Foundation (ISF) under grant number 0603225041.
The research leading to these results has received funding from the European Research Council (ERC) under the European Union's Horizon 2020 research and innovation programme (grant agreement numbers 772225: MULTIPLES). 
DMB gratefully acknowledges support from UK Research and Innovation (UKRI) in the form of a Frontier Research grant under the UK government's ERC Horizon Europe funding guarantee (SYMPHONY; grant number: EP/Y031059/1), and a Royal Society University Research Fellowship (grant number: URF{\textbackslash}R1{\textbackslash}231631). 
GGT is supported by the German Deutsche Forschungsgemeinschaft (DFG) under Project-ID 496854903 (SA4064/2-1, PI Sander). 
AACS is funded by the Deutsche Forschungsgemeinschaft (DFG, German Research Foundation) in the form of an Emmy Noether Research Group -- Project-ID 445674056 (SA4064/1-1, PI Sander). 
GGT and AACS further acknowledges support from the Federal Ministry of Education and Research (BMBF) and the Baden-Württemberg Ministry of Science as part of the Excellence Strategy of the German Federal and State Governments.  
This paper benefited from discussions at the International Space Science Institute (ISSI) in Bern through ISSI International Team project 512 (Multiwavelength View on Massive Stars in the Era of Multimessenger Astronomy). 
DP acknowledges financial support by the Deutsches Zentrum f\"ur Luft und Raumfahrt (DLR) grant FKZ 50OR2005. 
JIV acknowledges support from the European Research Council for the ERC Advanced Grant 101054731.  
PAC is supported by the Science and Technology Facilities Council
research grant ST/V000853/1 (PI. V. Dhillon). 
JSV is supported by Science and Technology Facilities Council funding under grant number ST/V000233/1. 
DFR is thankful for the support of the CAPES-Br and FAPERJ/DSC-10 (SEI-260003/001630/2023).
This work has received funding from the European Research Council (ERC) under the European Union’s Horizon 2020 research and innovation programme (Grant agreement No.\ 945806) and is supported by the Deutsche Forschungsgemeinschaft (DFG, German Research Foundation) under Germany’s Excellence Strategy EXC 2181/1-390900948 (the Heidelberg STRUCTURES Excellence Cluster).
\end{acknowledgements}

\bibliographystyle{aa}
\bibliography{papers}

\begin{appendix}

\section{Cross-correlation errors}
\label{sec:errors}

Leaving aside the issue of whether or not the peak of the CCF of two spectra represents a true RV shift, the problem of determining the error in the shift reduces to that of determining the uncertainty of the position of the CCF maximum.
In this paper, and others in the series, we adopt the approach of \citet{zucker2003} who used a maximum-likelihood analysis to derive a convenient analytic expression for the error ($\sigma$) reproduced here as:
\begin{equation}
    \sigma^2 = - \left[ N \frac{C"(\hat{s})}{C(\hat{s})} \frac{C^2(\hat{s})}{1-C^2(\hat{s})} \right]^{-1} ,
\end{equation}
where $C$ and $C"$ are the values of the CCF and its second derivative at the peak position ($\hat{s}$), and $N$ is the number of pixels in the spectrum. 
Typically the s/n is high enough such that both  $C$ and $C"$ can be reliably determined using a simple quadratic fit to the 3-pixel window centered on the maximum of the CCF.
However in cases where the s/n of one or more epochs is below $\sim$30 the CCF contains structure that results in small sputious offsets in velocity. 
In those cases a simple 3-pixel median smoothing function, and 7-pixel window, is found to give superior results, and since the data are over-sampled by a factor of $\sim$3 there is no significant loss of resolution. This approach was adopted for sources BLOeM~3-085, 3-092, 3-102, 4-004, 4-072, 4-114, 5-086, 6-007, 6-097, and 8-091.

Simulations by \citet{zucker2003} for cool star spectra validated that this equation does indeed reproduce the measured errors, given their assumptions. 
However a simplifying assumption in their analysis is that the noise in the data is both Gaussian and independent of pixel position. Here we have investigated the impact of adopting a Poisson noise model for the spectra, and we have also used continuum dominated spectra, that are appropriate for the bulk of the BAF supergiants under consideration here.

\begin{figure}
    \centering
    \includegraphics[width=1.0\linewidth]{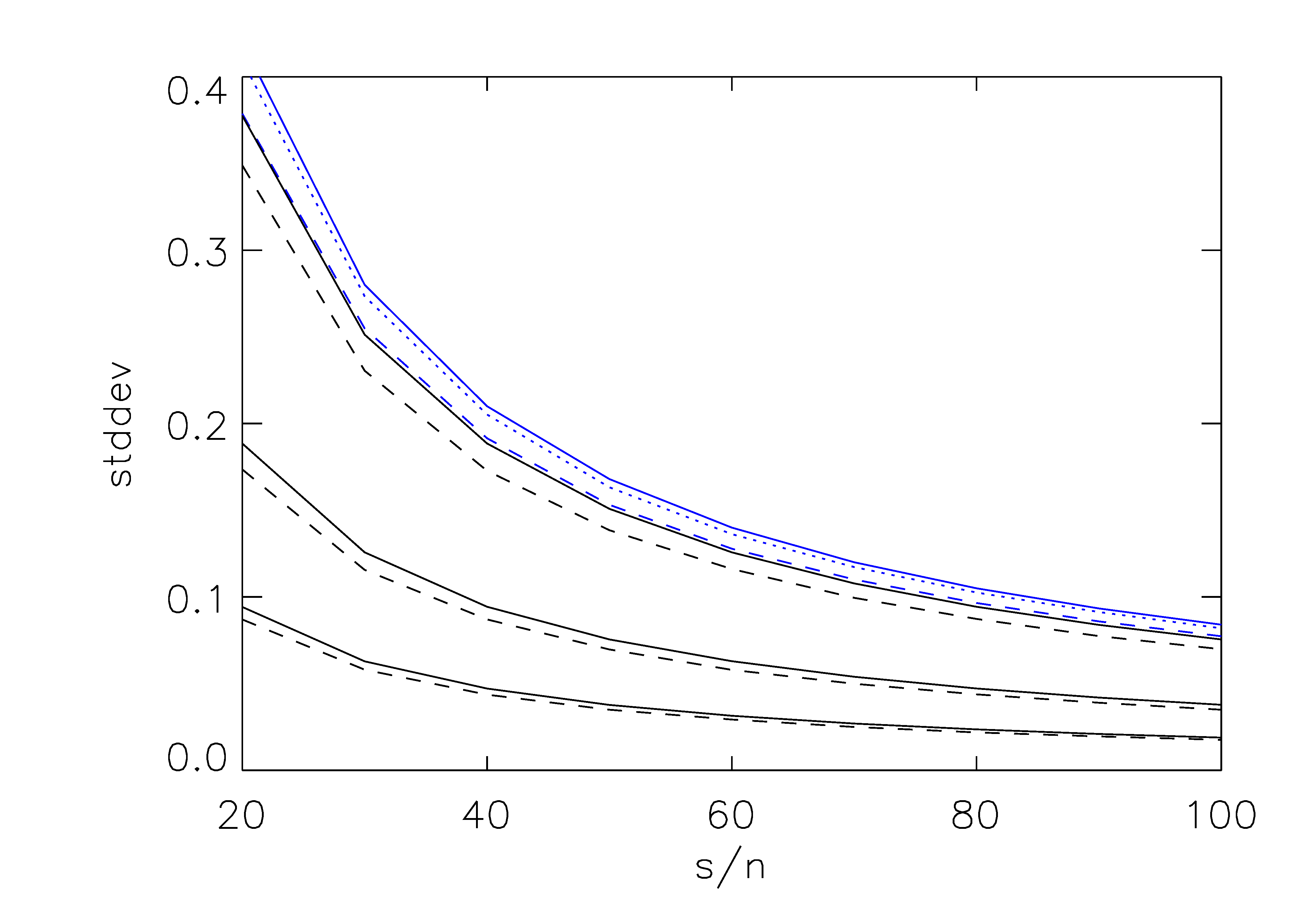}
    \caption{Dashed black lines -- plot of errors (stddev) of CCF positions versus signal-to-noise obtained in $10^5$ simulations of simple spectra consisting of 1, 4 and 16 absorption lines (ordering is top to bottom). Solid black lines -- expected trends due to equation A.1 which is larger by approximately 8\%. Note that the errors halve for each increase in the number of lines by a factor of 4. Blue lines are the equivalent results for a template A-supergiant spectrum, with the dotted line illustrating the explicit errors obtained from the simulation of Gaussian noise. The dashed blue line has been shifted upwards by 0.02 units for clarity.}
    \label{fig:sigma_s2n}
\end{figure}

In our first simulation we made three artificial spectra consisting of 1, 4 and 16 well separated non-overlapping Gaussian lines each with $\sigma=3$ units in a grid of uniform unit velocity spacing, and central depth 0.5 of the normalized continuum. We ran $10^5$ iterations of shifting each spectrum a fixed non-integer amount, adding Poisson noise, and using the CCF to estimate the spectrum shifts relative to the original template. 
The `observed' uncertainties were obtained by determining the standard deviation of the differences between the determined and actual shifts in the data, and are plotted in Figure \ref{fig:sigma_s2n}, as are the error estimates derived using equation A.1. 
We see that the Poisson noise errors are approximately 8\% smaller than predicted, and also that the errors decrease as the square root of the number of lines, rather than $N$, which is fixed in all simulations.
This is readily understood as the test spectrum is mostly continuum, unlike the case for cool stars considered in \citet{zucker2003}, which suggests a better description of $N$ to be the number of pixels that carry significant information.

\begin{figure}
    \centering
    \includegraphics[width=1.0\linewidth]{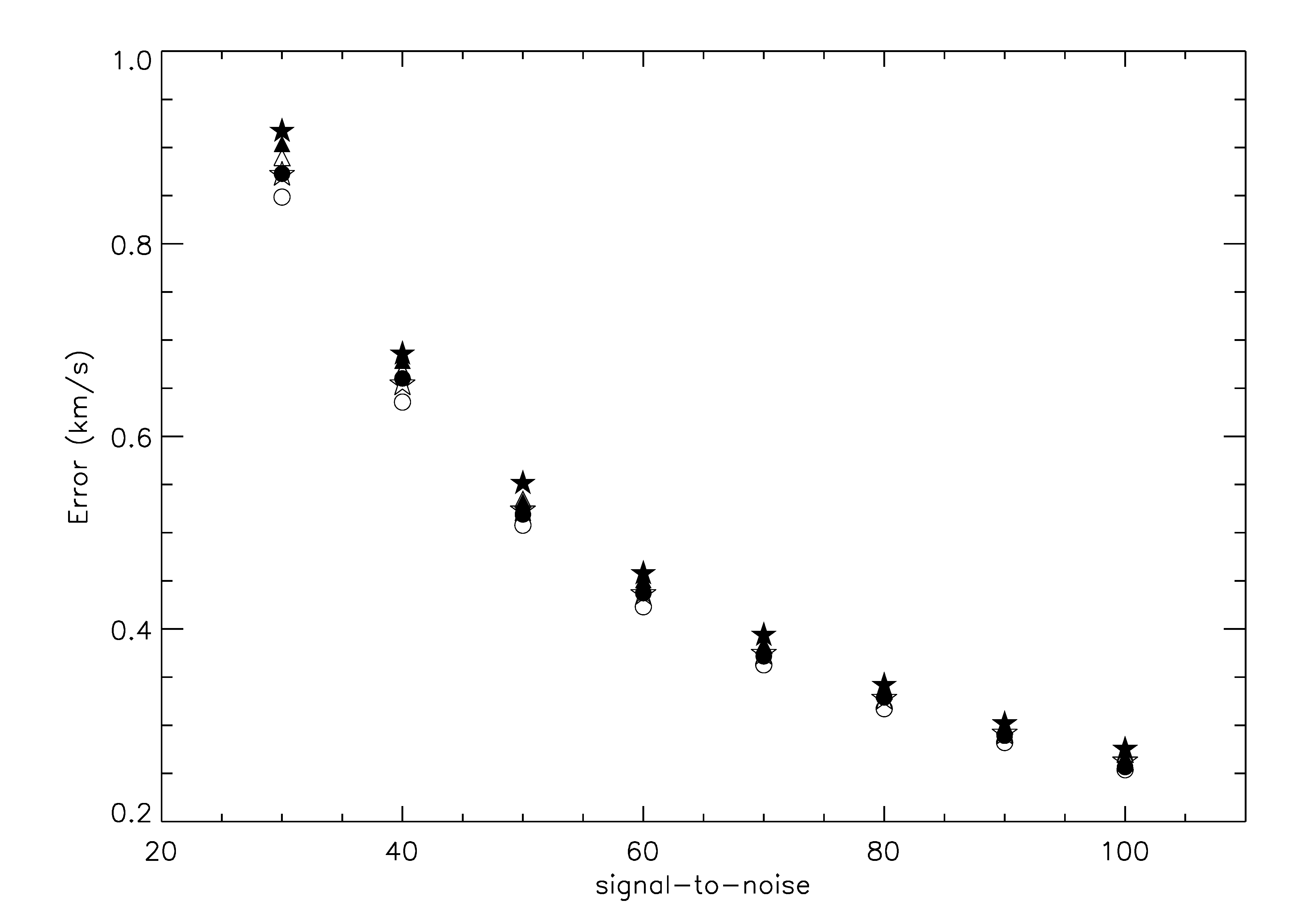}
    \caption{Plot of cross-correlation error versus signal-to-noise for a simulated BLOeM spectrum: Filled symbols indicate results from the Monte Carlo simulation, open symbols are results derived from equation A.1 modified as discussed in the test. Triangles, circles and stars denote resampled pixel sizes of 14, 7 and 1\',\kms respectively.}
    \label{fig:realsimulation}
\end{figure}

A more realistic second simulation adopted a CMFGEN A-supergiant spectrum with T$_{\rm eff}$ = 9750\,K, $\log$\,g = 2.0 and SMC  metallicity as the template spectrum, which has a range of line strengths in the relevant cross-correlation window.
This spectrum, as in the above example, is sampled at constant velocity intervals of 1\,\kms in order compare with the rather simple simulations discussed above. 
These results are also shown in Figure \ref{fig:sigma_s2n}, as are the explicit results for the calculated errors assuming Gaussian noise. 
We see that equation A.1 overestimates the actual errors by about 11\% in the case of this template. Similar results are obtained for templates with other representative stellar parameters.

However the situation for the BLOeM data is a little more complicated as the original data are re-binned from the 4000 native wavelength calibrated pixels to a uniform 0.2\AA\ pixel size, or 3056 pixels. 
For cross-correlation in velocity space these spectra must be resampled to a constant velocity grid of pixels (or logarithmic spacing in wavelength). 
Typically one might choose a velocity interval that is similar to the original pixel size, in this case $\sim$14\,\kms, however there may be value in sub-sampling pixels, for example to improve resolution of a final co-added data product.  
Assuming resampling simply interpolates onto a new wavelength grid, as is the case here, the consideration of equation A.1 tells us that one might expect the errors from this equation need to multiplied by a factor $\sqrt{\rm{old\_pixel\_size/new\_pixel\_size}}$.

This supposition was esseentially confirmed with Monte-Carlo simulations of cross-correlating an artificial BLOeM spectrum with a range of Poisson noise levels against the original noiseless spectrum. The artificial data were generated from a high resolution CMFGEN spectrum (as above), convolved with FLAMES resolution, and rebinned to 3056 0.2\,\AA\ pixels before resampling to a uniform velocity scale.
The formal errors from the cross-correlations were calculated as noted above, and the observed errors were determined from 10000 simulations, the formal errors being taken as the mean of the 10000 results individual results.
The results are shown in Figure\,\ref{fig:realsimulation} confirming that the above procedure for computing the errors in the case of resampling to a smaller pixel size, which is in good agreement with our simulations. We have repeated this Monte Carlo approach for the procedure adopted to measure radial velocities, and outlined in section \ref{sec:methods}, essentially finding the same result. Namely, the errors inferred from our modified error estimate agree with those derived from our Monte Carlo simulations.

However, since the likely errors due to systematic effects as described in Appendix \ref{sec:rv_corr} likely dominate the error budget for the BAF supergiants we refrain from correcting the formal errors here, and will return to this issue when future data releases are available.

\section{Radial velocity corrections}
\label{sec:rv_corr}

\begin{figure*}
    \centering
    \includegraphics[width=0.95\linewidth]{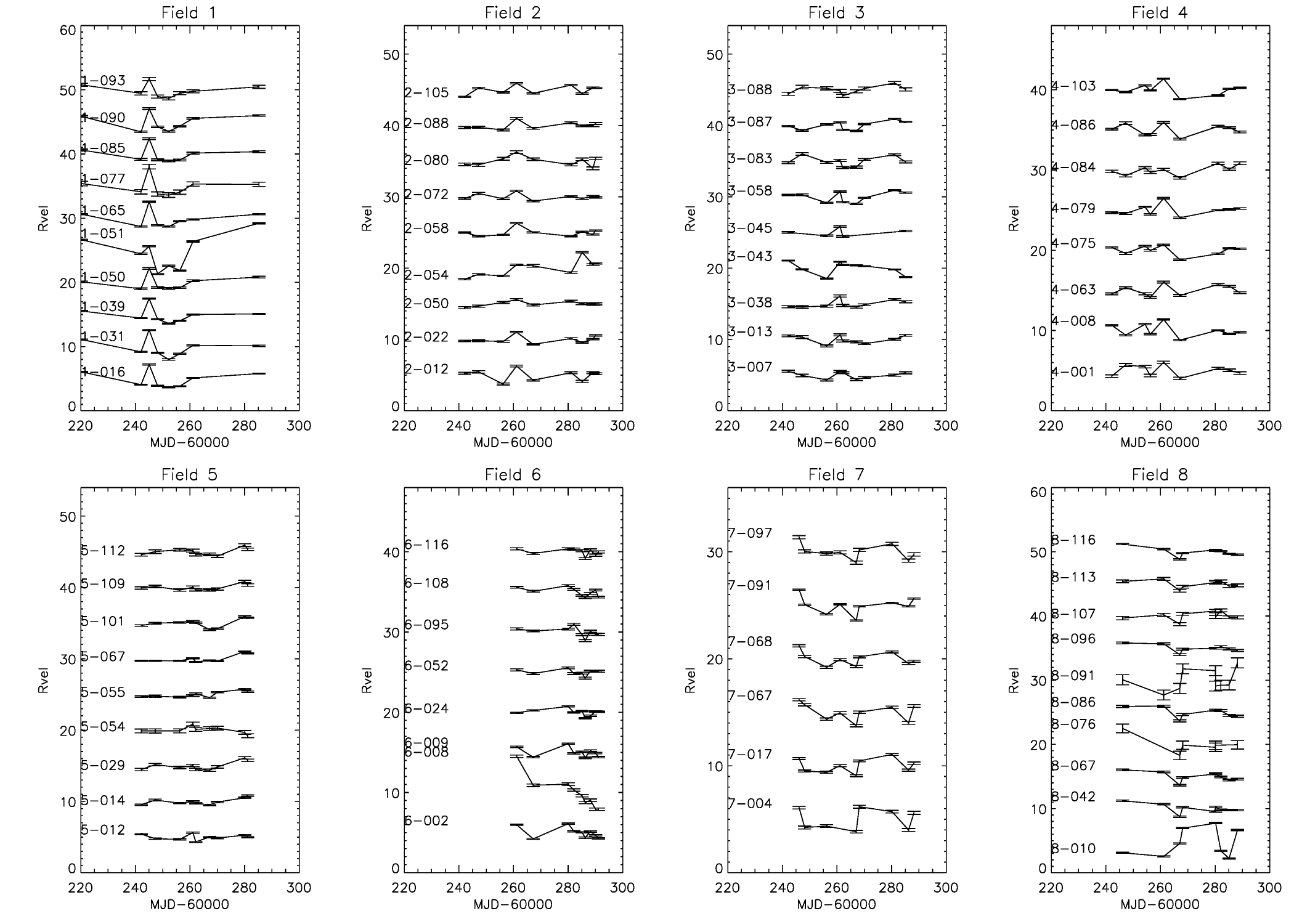}
    \caption{For each BLOeM field, noted in the heading of each panel, are plotted the measured radial velocities of the A-type supergiants. For ease of comparison, the median value has been subtracted from each source, and multiples of 5\,\kms\ added to separate the resulting trends. Note the correlated behaviour across many sources in each field.}
    \label{fig:perfield}
\end{figure*}

\begin{figure*}
    \centering
    \includegraphics[width=0.95\linewidth]{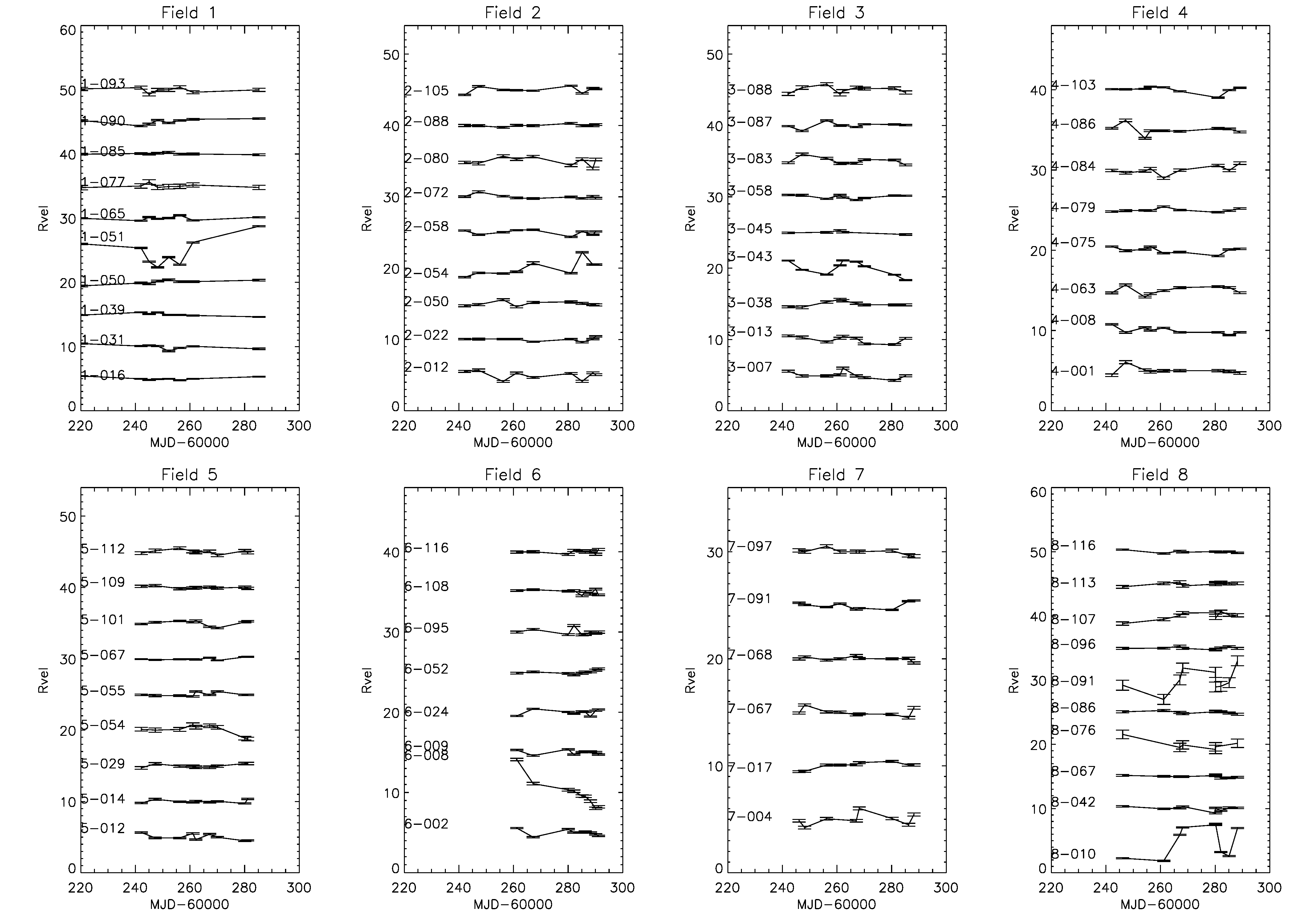}
    \caption{As for Figure\,\ref{fig:perfield} but with the corrections as tabulated in Table\,\ref{tab:corrections}.}
    \label{fig:perfield_corrected}
\end{figure*}

\begin{table*}[]
\centering
\caption{Velocity corrections (in \kms) for each field and epoch that have to be added to the measured radial velocities in Table\,\ref{tab:rvels} to mitigate systematic offsets in the data as discussed in section \ref{sec:methods}.}
\begin{tabular}{crrrrrrrrr} \hline
Field & \multicolumn{9}{c}{Epoch Number} \\
No.   & 1 & 2 & 3 & 4 & 5 & 6 & 7 & 8 & 9 \\ \hline
1 &  -0.45 &  -0.58  &  0.97 &  -2.34  &  1.10  &  1.39  &  0.98 &  -0.08 &  -0.41 \\
2 &   0.24 &   0.19  &  0.33 &  -1.02  &  0.34  & -0.12  &  0.00 &  -0.05 &  -0.17 \\
3 &   0.10 &   0.04  &  0.68 &  -0.37  &  0.76  &  0.64  &  0.08 &  -0.64 &  -0.32 \\
4 &   0.19 &   0.42  & -0.34 &   0.54  & -0.99  &  1.04  & -0.19 &  -0.08 &   0.07 \\
5 &   0.12 &   0.00  &  0.09 &  -0.22  &  0.20  &  0.29  &  0.00 &  -0.91 &  -0.57 \\
6 &  -0.37 &   0.23  & -0.67 &  -0.13  &  0.00  &  0.80  & -0.08 &   0.14 &   0.33 \\
7 &  -1.29 &  -0.09  &  0.61 &   0.00  &  0.95  & -0.23  & -0.72 &   0.42 &  -0.23 \\
8 &  -0.82 &  -0.66  &  1.36 &   0.16  & -0.23  & -0.12  & -0.15 &   0.39 &   0.35 \\ \hline
\end{tabular}
\label{tab:corrections}
\end{table*}

As was the case for the analysis of the cool, red supergiants from the VFTS survey~\citep{Patrick2019}, the FLAMES fields taken on different nights exhibit small temporal drifts in the wavelength calibration, which manifest as an offset in measured radial velocities.
This is most noticeable for the AF-type supergiants that enable high precision measurements due to the number and sharpness of their metal lines. 
Figure~\ref{fig:perfield} shows the temporal trends of their radial velocities for each field, from which it is evident that there are correlations between different stars in each field. This is clearly an artifact, most likely due to small thermal drifts affecting the wavelength scale, a known issue for the FLAMES spectrograph.

For each field we define a subset of sources that have small dispersion in their mean values, subtract their median velocities, determine the median offset at each epoch across all sources, and correct the observed velocities by this amount. The process is then repeated to convergence, typically only 2--3 iterations per field being required. 
Table\,\ref{tab:corrections} lists the resulting RV corrections that should be applied to observations in the field at a given epoch, while Figure\,\ref{fig:perfield_corrected} illustrates the impact of their application.
Clearly there are still a few issues with some stars. For example, in Field 8, stars 8-010 and 8-091 still display an apparent correlation. However, the full two years of BLOeM data should permit further refinement of this approach.






\onecolumn

\begin{centering}
\fontsize{9.pt}{10pt}\selectfont 
\setlength{\tabcolsep}{2pt}
\begin{table*}[ht]
\small
\caption{Measured radial velocities (\kms) and uncertainties for each epoch, as numbered. A value of zero indicates there are no data for that epoch number.}
\label{tab:rvels}
\centering
\begin{tabular}{lccc}
\hline
BLOeM & MJD & RV & error \\
\hline

1-001 & 60219.105 & 148.15 & 0.124 \\
1-001 & 60220.259 & 149.022 & 0.149 \\
1-001 & 60242.054 & 146.746 & 0.133 \\
1-001 & 60245.022 & 149.213 & 0.15 \\
1-001 & 60248.149 & 146.21 & 0.124 \\
1-001 & 60252.212 & 143.295 & 0.246 \\
1-001 & 60256.289 & 147.009 & 0.125 \\
1-001 & 60261.05 & 147.414 & 0.172 \\
1-001 & 60285.199 & 147.765 & 0.131 \\
1-008 & 60219.105 & 169.551 & 0.072 \\
1-008 & 60220.259 & 169.709 & 0.085 \\
1-008 & 60242.054 & 168.212 & 0.078 \\
1-008 & 60245.022 & 171.325 & 0.091 \\
1-008 & 60248.149 & 167.971 & 0.074 \\
1-008 & 60252.212 & 167.833 & 0.115 \\

\hline
\end{tabular}
\end{table*}
\end{centering}

\twocolumn

\onecolumn

\begin{centering}
\begin{center}
\begin{longtable}{lllllllllrll} 
    \caption{Summary table listing mean radial velocities, their standard deviation ($\sigma$), median value, and peak-to-peak velocity (p2p). Also listed is the median of individual errors, the number of epochs available, and the number of pairs of measurements satisfying equation 1. The effective temperature and logarithmic luminosity are those derived here and used in construction of the HRD. Candidate spectroscopic binaries are flagged as `$var$' according to Figure\,\ref{fig:BAFhist} and `LP' marks the potential long period system that does not meet the binary threshold. Candidate $\alpha$ Cyg variables are also indicated.} \\ \hline
BLOeM id & Sp Type  & mean & $\sigma$ & median &  p2p  & median & no. & n$_4$ & T$_{eff}$ & log\,L & Comment\\ 
&   & \kms & \kms & \kms & \kms  & error\,\kms & epochs &  & K & L$_{\odot}$  & \\ \hline
  1-001 &  B9 Iab   & 147.27 & 1.07 &  147.35 & 3.76   & 0.50	& 9  & 0      & 10250	& 4.27 &\\ 
  1-008 &  B9 Iab   & 169.11 & 0.17 &  169.10 & 0.63   & 0.29	& 9  & 0      & 10750	& 4.63 &\\ 
  1-016 &  A2 Ib    & 116.99 & 0.22 &  116.97 & 0.66   & 0.28	& 9  & 0      & 8250	& 4.70 &\\ 
  1-028 &  B8 Iab/Ia& 177.38 & 1.57 &  176.98 & 4.41   & 0.45	& 9  & 7      & 12500	& 5.01 &\\ 
  1-031 &  A0 Iab   & 156.42 & 0.37 &  156.50 & 1.15   & 0.34	& 9  & 0      & 9500	& 4.65 &\\ 
  1-039 &  A2 II/Ib & 144.27 & 0.22 &  144.19 & 0.71   & 0.29	& 9  & 0      & 7750	& 4.40 &\\ 
  1-050 &  A7 Ib    & 134.06 & 0.31 &  134.15 & 0.93   & 0.49	& 9  & 0      & 7500	& 4.15 &\\ 
  1-051 &  A0 Ia    & 149.65 & 2.09 &  150.01 & 6.39   & 0.36	& 9  & 23     & 9500	& 5.33 & $\alpha$ Cyg $var$\\ 
  1-062 &  B8 Iab   & 150.87 & 0.46 &  150.95 & 1.31   & 0.35	& 9  & 0      & 13000	& 4.73 &\\ 
  1-064 &  B9 Ib    & 191.52 & 0.75 &  191.76 & 2.12   & 0.87	& 9  & 0      & 9240	& 3.76 &\\ 
  1-065 &  A7 Ib    & 176.77 & 0.26 &  176.79 & 0.86   & 0.36	& 9  & 0      & 7500	& 4.34 &\\ 
  1-077 &  A2 II/Ib & 141.84 & 0.27 &  141.79 & 0.88   & 1.28	& 9  & 0      & 8500	& 4.31 &\\ 
  1-085 &  A7 Ib/ab & 119.58 & 0.14 &  119.55 & 0.50   & 0.56	& 9  & 0      & 7500	& 4.19 &\\ 
  1-090 &  A1 Ib    & 130.46 & 0.42 &  130.68 & 1.15   & 0.42	& 9  & 0      & 8740	& 3.80 &\\ 
  1-093 &  A7 Ib    & 148.83 & 0.36 &  148.82 & 1.10   & 0.91	& 9  & 0      & 7500	& 4.42 &\\ 
  1-112 &  B9 Ia    & 193.62 & 2.95 &  193.60 & 10.54  & 0.53	& 8  & 14     & 11500	& 5.34 & $\alpha$ Cyg $var$\\ 
  1-114 & F5:	    & 174.15 & 1.42 &  173.53 & 3.85   & 0.65	& 9  & 2      & 7000	& 5.10 & LP\\ 
  2-012 &  A0 Ib    & 105.86 & 0.58 &  106.11 & 1.57   & 0.51	& 9  & 0      & 9250	& 4.00 &\\ 
  2-022 &  A0 Ib    & 98.34  & 0.24 &  98.38  & 0.82   & 0.37	& 9  & 0      & 9500	& 4.23 &\\ 
  2-050 &  A5 Ib    & 104.22 & 0.30 &  104.12 & 0.96   & 0.54	& 9  & 0      & 7500	& 4.10 &\\ 
  2-054 &  A0 Ia    & 152.07 & 1.08 &  151.56 & 3.52   & 0.38	& 9  & 5      & 10000	& 4.87 &\\ 
  2-058 &  A5 Ib    & 130.32 & 0.34 &  130.45 & 1.0    & 0.34	& 9  & 0      & 8250	& 4.42 &\\ 
  2-065 & F5:	    & 148.81 & 0.28 &  148.79 & 0.91   & 0.56	& 9  & 0      & 6500	& 5.04 &\\ 
  2-067 &  B8 Ib    & 105.75 & 0.73 &  105.92 & 2.39   & 0.34	& 9  & 1      & 13000	& 4.63 &\\ 
  2-068 &  B9 Iab   & 154.03 & 0.24 &  154.06 & 0.77   & 0.45	& 9  & 0      & 10000	& 4.53 &\\ 
  2-072 &  A2 II/Ib & 115.24 & 0.29 &  115.21 & 0.93   & 0.42	& 9  & 0      & 8990	& 4.06 &\\ 
  2-073 &  B9 Ia    & 164.18 & 2.03 &  165.51 & 4.45   & 0.37	& 4  & 3      & 11490	& 4.77 &\\ 
  2-080 &  A2 II/Ib & 136.05 & 0.57 &  136.27 & 1.71   & 0.69	& 9  & 0      & 8500	& 4.17 &\\ 
  2-088 &  A5 Ib    & 128.51 & 0.15 &  128.47 & 0.57   & 0.49	& 9  & 0      & 7500	& 4.01 &\\ 
  2-092 &  B8 Iab   & 157.85 & 1.88 &  158.02 & 6.12   & 0.38	& 9  & 17     & 11990	& 4.97 & $\alpha$ Cyg $var$\\ 
  2-093 &  B8 Ia    & 150.72 & 3.07 &  150.52 & 9.57   & 0.58	& 9  & 16     & 13000	& 5.62 & $\alpha$ Cyg $var$\\ 
  2-101 &  B9 Ib    & 160.20 & 0.44 &  160.24 & 1.19   & 0.62	& 8  & 0      & 9500	& 4.19 &\\ 
  2-105 &  A2 II/Ib & 137.40 & 0.42 &  137.39 & 1.28   & 0.37	& 9  & 0      & 8740	& 4.03 &\\ 
  2-108 &  B5 II    & 140.81 & 0.85 &  141.12 & 2.19   & 0.71	& 9  & 0      & 12500	& 3.94 &\\ 
  2-115 &  B5 Ib    & 150.37 & 0.68 &  150.55 & 1.94   & 0.50	& 9  & 0      & 14000	& 4.53 &\\ 
  3-007 &  A5 Ib    & 144.24 & 0.50 &  144.13 & 1.72   & 0.55	& 9  & 0      & 7750	& 4.05 &\\ 
  3-009 &  B5 II    & 137.23 & 1.39 &  137.39 & 4.60   & 1.06	& 9  & 0      & 14500	& 4.30 &\\ 
  3-013 &  A0 Ib    & 121.07 & 0.45 &  121.23 & 1.23   & 0.50	& 9  & 0      & 9250	& 3.95 &\\ 
  3-024 &  B5 II    & 123.10 & 1.42 &  122.98 & 4.87   & 0.75	& 9  & 0      & 12500	& 3.93 &\\ 
  3-038 &  A0 Ia    & 142.96 & 0.37 &  142.81 & 1.11   & 0.58	& 9  & 0      & 9500	& 4.27 &\\ 
  3-041 &  B8 Ib    & 146.93 & 0.67 &  146.81 & 2.27   & 0.47	& 9  & 0      & 14000	& 4.67 &\\ 
  3-043 &  A2 Iab   & 122.43 & 0.99 &  122.70 & 2.76   & 0.26	& 9  & 11     & 8250	& 4.71 &\\ 
  3-045 &  A0 Ib    & 105.20 & 0.21 &  105.23 & 0.57   & 0.43	& 5  & 0      & 9500	& 4.15 &\\ 
  3-058 &  A0 Iab   & 159.11 & 0.26 &  159.27 & 0.71   & 0.34	& 9  & 0      & 9500	& 4.14 &\\ 
  3-068 &  B5 Ib    & 150.55 & 0.41 &  150.60 & 1.08   & 0.48	& 9  & 0      & 14500	& 4.53 &\\ 
  3-083 &  A0 Ib    & 131.49 & 0.47 &  131.26 & 1.50   & 0.50	& 9  & 0      & 10000	& 4.08 &\\ 
  3-085 &  B9 Iab   & 146.45 & 2.67 &  145.45 & 7.43   & 3.36	& 9  & 0      & 9500	& 4.13 &\\ 
  3-087 &  A2 II/Ib & 151.38 & 0.39 &  151.42 & 1.47   & 0.34	& 9  & 0      & 8250	& 4.25 &\\ 
  3-088 &  A2 II    & 130.92 & 0.47 &  131.08 & 1.41   & 0.81	& 9  & 0      & 8500	& 4.06 &\\ 
  3-092 &  B5 II    & 160.97 & 2.00 &  161.27 & 6.33   & 0.79	& 9  & 4      & 14500	& 4.20 & $var$\\ 
  3-099 &  B9 Iab   & 152.25 & 0.47 &  152.25 & 1.61   & 0.41	& 9  & 0      & 11000	& 4.40 &\\ 
  3-102 &  B8 II-Ib & 163.60 & 2.67 &  163.00 & 6.83   & 1.02	& 9  & 6      & 12500	& 4.04 & $var$\\ 
  3-104 &  B8 Iab   & 148.89 & 1.21 &  148.60 & 3.47   & 0.48	& 9  & 2      & 14000	& 4.61 &\\ 
  3-106 &  B5 Ib    & 130.31 & 1.65 &  130.94 & 4.70   & 0.44	& 9  & 10     & 14500	& 4.64 & LP\\ 
  3-110 &  B8 II-Ib & 143.45 & 1.22 &  143.24 & 3.63   & 0.62	& 9  & 2      & 12500	& 4.12 &\\ 
  3-114 & F2:	    & 123.48 & 0.41 &  123.37 & 0.94   & 0.45	& 9  & 0      & 6250	& 4.08 &\\ 
  4-001 &  A1 Ib    & 144.29 & 0.44 &  144.27 & 1.62   & 0.58	& 9  & 0      & 9250	& 4.32 &\\ 
  4-004 &  B5 II:   & 148.53 & 4.03 &  148.47 & 12.87  & 1.78	& 9  & 3      & 12500	& 3.77 & $var$\\ 
  4-006 & F2:	    & 177.46 & 0.26 &  177.45 & 0.93   & 0.46	& 9  & 0      & 7000	& 4.79 & UV excess\\ 
  4-008 &  A0 Ib    & 181.55 & 0.42 &  181.32 & 1.35   & 0.30	& 9  & 0      & 9750	& 4.40 &\\ 
  4-054 &  B9 Ib    & 153.27 & 0.45 &  153.23 & 1.27   & 0.85	& 9  & 0      & 12500	& 4.22 &\\ 
  4-060 &  B8 II-Ib & 127.46 & 0.76 &  127.87 & 1.91   & 0.73	& 9  & 0      & 12500	& 4.11 &\\ 
  4-063 &  A7 Ib    & 149.10 & 0.50 &  149.08 & 1.51   & 0.44	& 9  & 0      & 7500	& 4.44 &\\ 
  4-072 &  B9 Ia    & 169.13 & 1.86 &  169.47 & 5.50   & 0.34	& 9  & 19     & 11000	& 4.76 & $\alpha$ Cyg $var$\\ 
  4-075 &  A0 Ib    & 148.05 & 0.38 &  148.16 & 1.17   & 0.34	& 9  & 0      & 9750	& 4.32 &\\ 
  4-079 &  A0 Ib    & 151.37 & 0.21 &  151.32 & 0.70   & 0.34	& 9  & 0      & 9500	& 4.33 &\\ 
  4-084 &  A0 Iab   & 135.36 & 0.53 &  135.31 & 1.86   & 0.55	& 9  & 0      & 9750	& 4.56 &\\ 
  4-086 &  A2 II/Iab& 156.58 & 0.58 &  156.47 & 2.24   & 0.43	& 9  & 0      & 8740	& 3.92 &\\ 
  4-091 &  B8 Ib/Iab& 143.31 & 0.42 &  143.24 & 1.37   & 0.59	& 9  & 0      & 12500	& 4.57 &\\ 
  4-103 &  A5 Ib    & 179.22 & 0.41 &  179.29 & 1.35   & 0.27	& 9  & 0      & 7750	& 4.45 &\\ 
  4-112 &  B5 II    & 154.36 & 1.03 &  154.25 & 3.65   & 0.81	& 9  & 0      & 14500	& 4.15 &\\ 
  4-114 &  B5 II    & 189.96 & 2.13 &  189.82 & 6.96   & 1.35	& 9  & 0      & 14500	& 4.16 &\\ 
  5-005 & F2:	    & 136.56 & 6.98 &  132.37 & 16.06  & 0.47	& 7  & 11     & 6250	& 3.87 & Cepheid\\ 
  5-012 &  A2 II/Ib & 165.38 & 0.43 &  165.26 & 1.17   & 0.37	& 9  & 0      & 8500	& 4.12 &\\ 
  5-014 &  A0 Ib    & 166.02 & 0.22 &  165.98 & 0.65   & 0.36	& 9  & 0      & 9500	& 4.43 &\\ 
  5-029 &  A7 Ib    & 168.46 & 0.24 &  168.41 & 0.62   & 0.65	& 9  & 0      & 7500	& 4.30 &\\ 
  5-036 &  B9 Ib    & 188.06 & 1.23 &  188.05 & 3.96   & 1.04	& 9  & 0      & 9240	& 3.81 &\\ 
  5-052 &  B9 Iab   & 121.88 & 0.67 &  121.82 & 1.74   & 0.41	& 9  & 0      & 10250	& 4.47 &\\ 
  5-054 &  A5 Ib    & 156.73 & 0.72 &  156.84 & 2.0    & 0.98	& 9  & 0      & 8500	& 4.08 &\\ 
  5-055 &  A1 II/Ib & 168.54 & 0.25 &  168.48 & 0.71   & 0.38	& 9  & 0      & 8990	& 3.94 &\\ 
  5-067 &  A0 Iab   & 140.17 & 0.18 &  140.11 & 0.50   & 0.25	& 9  & 0      & 9750	& 4.72 &\\ 
  5-086 &  B9 Ib    & 191.16 & 3.37 &  191.69 & 12.02   & 1.96	& 9  & 1      & 9240	& 3.81 & $var$\\ 
  5-091 &  B8 Ib    & 171.91 & 0.60 &  172.0  & 1.92   & 0.36	& 9  & 0      & 13000	& 4.74 &\\ 
  5-098 &  B9 Ia    & 173.80 & 0.45 &  173.85 & 1.12   & 0.33	& 9  & 0      & 11500	& 4.59 &\\ 
  5-101 &  A5 Ib    & 138.39 & 0.37 &  138.52 & 1.04   & 0.39	& 9  & 0      & 8250	& 4.21 &\\ 
  5-102 & F0:	    & 191.72 & 0.39 &  191.72 & 1.49   & 0.50	& 9  & 0      & 7000	& 4.00 &\\ 
  5-109 &  A7 Iab   & 143.30 & 0.14 &  143.27 & 0.43   & 0.63	& 9  & 0      & 7500	& 4.27 & UV excess\\ 
  5-112 &  A0 Ib    & 188.94 & 0.27 &  188.99 & 0.99   & 0.71	& 9  & 0      & 9250	& 3.88 &\\ 
  6-002 &  A2 Ib    & 183.89 & 0.36 &  183.90 & 1.14   & 0.32	& 9  & 0      & 8740	& 4.42 &\\ 
  6-006 & F2:	    & 174.07 & 1.65 &  173.61 & 5.08   & 0.84	& 9  & 1      & 6500	& 4.45 & $var$\\ 
  6-007 &  B5 II    & 179.59 & 3.90 &  179.19 & 10.70  & 1.71	& 9  & 4      & 14500	& 4.03 & $var$\\ 
  6-008 &  A2 Ia    & 174.88 & 1.84 &  174.44 & 6.09   & 0.59	& 9  & 7      & 8740	& 5.57 & $var$ UV excess\\ 
  6-009 &  A2 II/Ib & 160.83 & 0.27 &  160.87 & 0.78   & 0.33	& 9  & 0      & 8250	& 4.26 &\\ 
  6-015 & F2:	    & 171.47 & 0.39 &  171.27 & 1.17   & 0.36	& 9  & 0      & 7000	& 4.29 &\\ 
  6-024 &  A5 Iab   & 173.86 & 0.34 &  173.90 & 0.97   & 0.29	& 9  & 0      & 8250	& 4.95 &\\ 
  6-052 &  A2 II/Ib & 169.73 & 0.23 &  169.76 & 0.75   & 0.44	& 9  & 0      & 8250	& 4.43 & UV excess\\ 
  6-095 &  A0 Ib    & 173.28 & 0.37 &  173.27 & 1.17   & 0.45	& 9  & 0      & 9250	& 4.03 &\\ 
  6-097 &  B8 II-Ib & 172.06 & 4.30 &  173.62 & 13.56  & 1.86	& 9  & 5      & 12500	& 3.95 & $var$\\ 
  6-108 &  A0 Iab   & 170.10 & 0.30 &  170.19 & 0.83   & 0.40	& 9  & 0      & 10000	& 4.41 &\\ 
  6-116 &  A7 Iab   & 170.75 & 0.21 &  170.75 & 0.60   & 0.50	& 9  & 0      & 7500	& 4.07 &\\ 
  7-004 &  A0 Iab   & 119.48 & 0.55 &  119.53 & 1.80   & 0.49	& 8  & 0      & 9750	& 4.20 &\\ 
  7-012 &  B5 II    & 154.57 & 0.70 &  154.62 & 2.51   & 0.52	& 9  & 0      & 14500	& 4.19 &\\ 
  7-017 &  A2 Ib    & 149.71 & 0.33 &  149.78 & 0.97   & 0.34	& 9  & 0      & 8990	& 4.14 &\\ 
  7-046 &  B5 II    & 152.71 & 1.30 &  152.88 & 4.12   & 0.82	& 9  & 0      & 12500	& 4.27 &\\ 
  7-067 &  A0 Ib    & 139.94 & 0.36 &  139.88 & 1.19   & 0.40	& 9  & 0      & 9500	& 4.22 &\\ 
  7-068 &  A5 Ib    & 155.25 & 0.19 &  155.24 & 0.70   & 0.37	& 9  & 0      & 8250	& 4.34 &\\ 
  7-075 &  B8 Ib    & 156.24 & 0.54 &  156.01 & 1.61   & 0.57	& 9  & 0      & 12500	& 4.70 &\\ 
  7-091 &  A5 Iab   & 152.03 & 0.33 &  152.06 & 0.87   & 0.22	& 9  & 0      & 7500	& 4.95 &\\ 
  7-097 &  A7 Ib    & 165.27 & 0.27 &  165.28 & 0.93   & 0.57	& 9  & 0      & 7500	& 4.00 &\\ 
  7-106 &  B8 Ib    & 149.73 & 0.58 &  149.52 & 1.59   & 1.02	& 9  & 0      & 12500	& 4.70 &\\ 
  8-001 &  B8 Ib    & 156.53 & 1.01 &  156.14 & 2.92   & 0.37	& 9  & 7      & 12000	& 4.69 & LP\\ 
  8-010 &  A0 Ia    & 163.74 & 2.46 &  164.64 & 5.78   & 0.34	& 9  & 20     & 9750	& 5.10 & $\alpha$ Cyg $var$\\ 
  8-042 &  A5 Ib    & 134.24 & 0.33 &  134.34 & 1.06   & 0.41	& 9  & 0      & 8250	& 4.07 &\\ 
  8-061 &  B9 Ib    & 112.07 & 0.67 &  112.05 & 2.12   & 1.12	& 9  & 0      & 9240	& 3.97 &\\ 
  8-067 &  A7 Ib    & 152.01 & 0.18 &  151.98 & 0.55   & 0.50	& 9  & 0      & 7500	& 4.40 &\\ 
  8-072 & F2:	    & 148.53 & 0.18 &  148.55 & 0.54   & 0.43	& 7  & 0      & 7000	& 4.58 &\\ 
  8-076 &  A5 Ib    & 137.12 & 0.83 &  137.03 & 2.34   & 2.46	& 6  & 0      & 8990	& 4.26 &\\ 
  8-077 &  B5 II    & 142.59 & 1.14 &  142.79 & 3.78   & 0.70	& 8  & 0      & 14500	& 4.28 &\\ 
  8-082 &  B9 Iab   & 153.20 & 0.38 &  153.21 & 1.23   & 0.33	& 8  & 0      & 10500	& 4.51 &\\ 
  8-083 & F0:	    & 140.51 & 0.20 &  140.43 & 0.65   & 0.34	& 9  & 0      & 7000	& 3.87 &\\ 
  8-086 &  A7 Ib    & 133.84 & 0.19 &  133.91 & 0.58   & 0.61	& 9  & 0      & 7500	& 4.32 &\\ 
  8-089 & F2:	    & 139.34 & 0.39 &  139.45 & 1.10   & 0.38	& 6  & 0      & 6500	& 4.03 &\\ 
  8-091 &  A0 II    & 140.15 & 2.08 &  140.23 & 6.83   & 2.83	& 9  & 0      & 8990	& 4.03 &\\ 
  8-096 &  A7 Ib    & 126.49 & 0.18 &  126.44 & 0.57   & 0.55	& 9  & 0      & 7500	& 4.08 &\\ 
  8-097 & F0:	    & 146.28 & 0.11 &  146.25 & 0.21   & 0.39	& 3  & 0      & 7000	& 3.93 &\\ 
  8-101 & F0:	    & 137.88 & 0.22 &  137.79 & 0.62   & 0.34	& 9  & 0      & 7000	& 4.05 &\\ 
  8-107 &  A0 Ib    & 137.07 & 0.58 &  137.18 & 1.86   & 0.88	& 9  & 0      & 9250	& 3.96 &\\ 
  8-113 &  A5 Ib    & 139.60 & 0.24 &  139.68 & 0.72   & 0.81	& 9  & 0      & 7500	& 4.15 &\\ 
  8-116 &  A2 Ib    & 154.51 & 0.18 &  154.54 & 0.61   & 0.39	& 9  & 0      & 9250	& 4.43 &\\ \hline
    \label{tab:summary}
\end{longtable}    
\end{center}

\end{centering}

\twocolumn

\section{Radial velocity plots of candidates variables}
\label{sec:rv_plots}

\begin{figure*}
    \centering
    \includegraphics[width=0.85\linewidth]{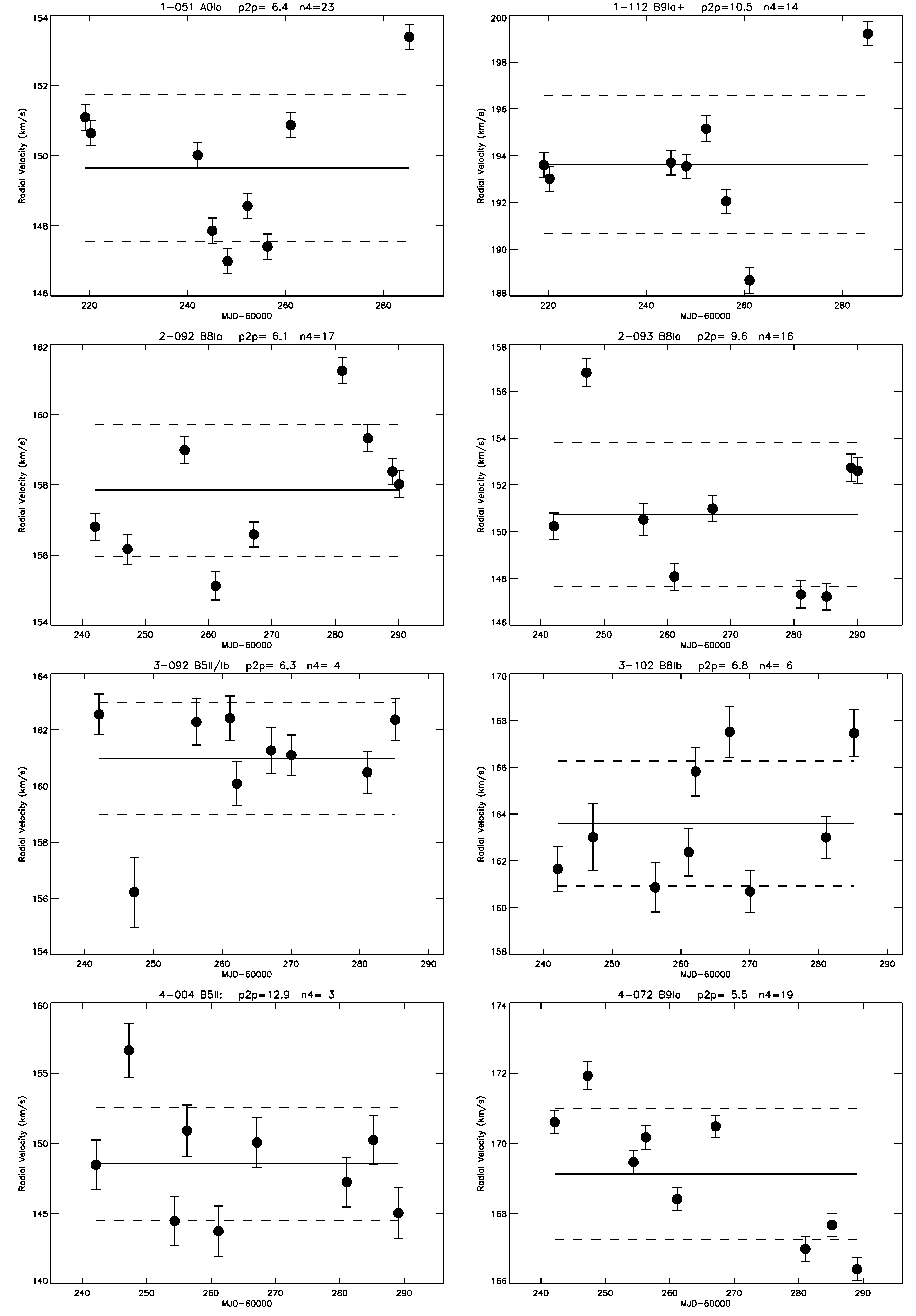}
    \caption{Radial velocity plots as function MJD for sources flagged as $var$ in Table \ref{tab:summary}, plot headers giving BLOeM identifiers, spectra types, peak-to-peak velocities, number of velocity pairs satisfying equation \ref{eq:sig}. The horizontal solid lines indicate the mean velocities, and the dashed lines the standard deviations. }
    \label{fig:plot_var_a}
\end{figure*}

\begin{figure*}
    \centering
    \includegraphics[width=0.85\linewidth]{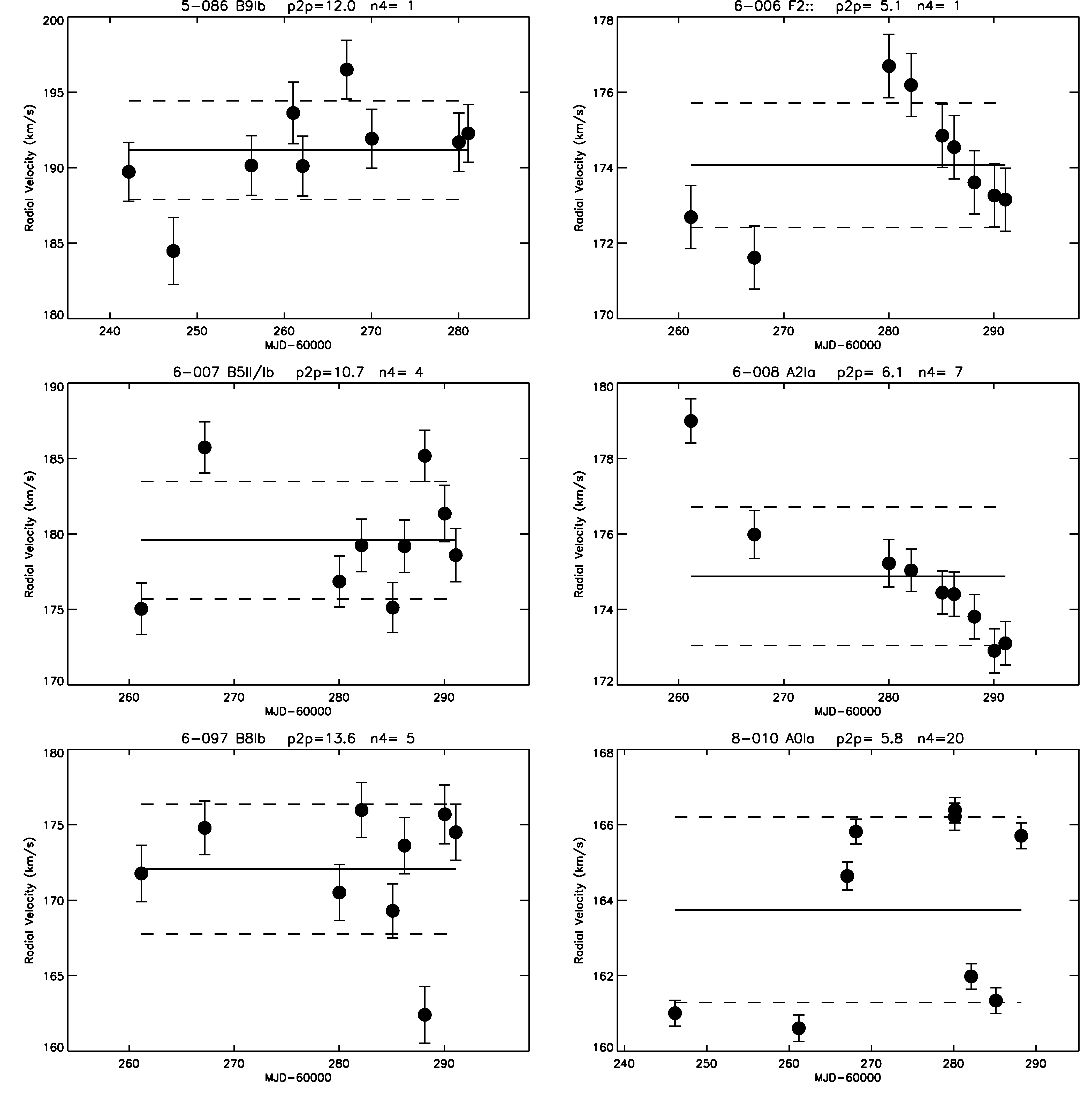}
    \caption{Additional sources as per Figure\,\ref{fig:plot_var_a}. }
    \label{fig:plot_var_b}
\end{figure*}

\begin{figure*}
    \centering
    \includegraphics[width=0.85\linewidth]{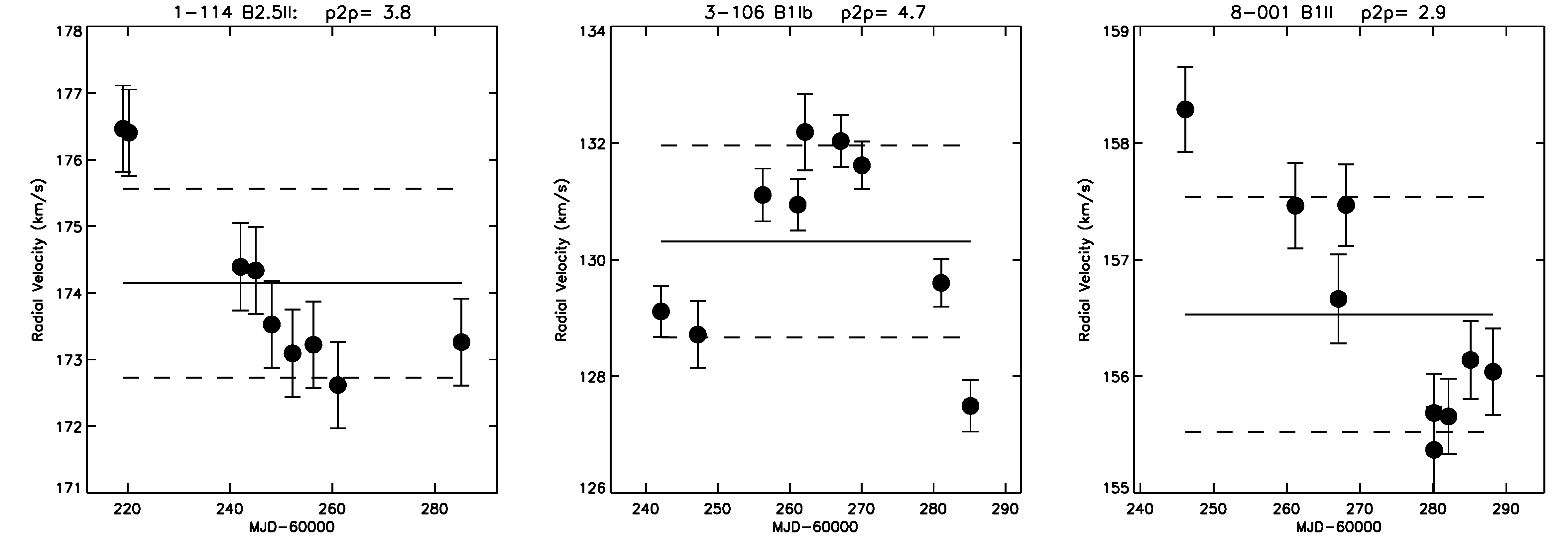}
    \caption{Radial velocity plots as function MJD for sources exhibiting possible binary motion, or perhaps long-term pulsation. Plot headers list BLOeM identifiers, spectra types, and peak-to-peak velocities \ref{eq:sig}. The horizontal solid lines indicate the mean velocities, and the dashed lines the standard deviations. }
    \label{fig:plot_other_var}
\end{figure*}

In this Appendix we plot the epoch RV data of all 14 sources satisfying Equations\,\ref{eq:rv_lim} and \ref{eq:sig}, Figures \ref{fig:plot_var_a} and \ref{fig:plot_var_b}, plus three examples of sources that do not pass the variability criteria yet exhibit signs of long-term correlated variability, Figure \ref{fig:plot_other_var}, that may indicate binarity or potentially (long-term) pulsations.

\end{appendix}

\end{document}